\documentclass{aa}

\makeatletter

\renewcommand*\aa@pageof{, page \thepage{} of \pageref*{LastPage}}

\usepackage{natbib}
\bibpunct{(}{)}{;}{a}{}{,}%
\def\bibfont{\aa@bibliographyfont}%
\providecommand{\@LN}[2]{}

\makeatother

\usepackage{graphicx}
\usepackage{newtxtext}
\usepackage[smallerops,cmbraces]{newtxmath}
\usepackage{xcolor}
\definecolor{xlinkcolor}{cmyk}{1,1,0,0}
\usepackage[
  bookmarks=true,         
  pdfnewwindow=true,      
  colorlinks=true,        
  linkcolor=xlinkcolor,   
  citecolor=xlinkcolor,   
  filecolor=xlinkcolor,   
  urlcolor=xlinkcolor,    
  final=true,
]{hyperref}

\usepackage{bm} 
\usepackage{textgreek} 

\usepackage[capitalize]{cleveref} 
\crefname{section}{Sect.}{Sects.}
\creflabelformat{equation}{#2#1#3} 
\crefname{enumi}{item}{items} 

\usepackage{siunitx} 
\DeclareSIUnit[number-unit-product = ]\percent{\char`\%} 

\usepackage[normalem]{ulem} 
\usepackage{csquotes} 
\usepackage{enumitem} 
\setlist{nosep} 
\definecolor{blackberry}{HTML}{8D1D75}
\definecolor{lightblue}{rgb}{0.1,0.5,0.89}
\definecolor{darkorange}{HTML}{FF8C00}

\newcommand*{\name}[1]{\textsc{#1}} 

\renewcommand{\vec}{\bm} 
\renewcommand{\tens}[1]{\bm{\mathrm{#1}}} 

\newcommand*{\bvalue}{$b$-value}
\newcommand*{\bvalues}{$b$-values}

\newcommand*{\LCDM}{\textLambda{}CDM}
\newcommand*{\gadget}[1]{\name{Gadget-#1}}
\newcommand*{\subfind}{\name{SubFind}}
\newcommand*{\wmap}{WMAP}
\newcommand*{\MagneticumBox}[2]{Box#1 (#2)}
\newcommand*{\sauron}{SAURON}
\newcommand*{\atlas}{ATLAS\textsuperscript{3D}}

\DeclareSIUnit\parsec{pc}
\DeclareSIUnit\dex{dex}
\DeclareSIUnit\h{\mathnormal{h}}
\DeclareSIUnit\year{yr}
\DeclareSIUnit\years{yrs}
\DeclareSIUnit\arcsec{arcsec}
\DeclareSIUnit\arcmin{arcmin}
\DeclareSIUnit\Msun{M_\odot}
\DeclareSIUnit\Rsun{R_\odot}
\DeclareSIUnit\Lsun{L_\odot}
\DeclareSIUnit\Rvir{\mathnormal{R}_\mathrm{vir}}
\DeclareSIUnit\Rhalf{\mathnormal{R}_{1/2}}
\DeclareSIUnit\Reff{\mathnormal{R}_e}
\DeclareSIUnit\erg{erg}
\DeclareSIUnit\angstrom{\text{Å}}

\newcommand*{\Msun}{\ensuremath{\mathrm{M}_\odot}} 
\newcommand*{\Rsun}{\ensuremath{\mathrm{R}_\odot}} 
\newcommand*{\Lsun}{\ensuremath{\mathrm{L}_\odot}} 
\newcommand*{\Rvir}{\ensuremath{R_\mathrm{vir}}} 
\newcommand*{\Rhalf}{\ensuremath{R_{1/2}}} 
\newcommand*{\Reff}{\ensuremath{R_e}} 
\newcommand*{\lambdaR}{\ensuremath{\lambda_R}} 
\newcommand*{\lambdaRhalf}{\ensuremath{\lambda_{R_{1/2}}}} 
\newcommand*{\jdmvec}{\ensuremath{\vec{j}_\mathrm{DM}}} 
\newcommand*{\jstvec}{\ensuremath{\vec{j}_*}} 

\begin{document}

\title{Galaxy shapes in Magneticum}
\subtitle{I. Connecting stellar and dark matter shapes to dynamical and morphological galaxy properties and the large-scale structure}
\titlerunning{Galaxy shapes in Magneticum I.}

\author{
    Lucas M. Valenzuela\inst{\ref{inst:usm}} 
    \and
    Rhea-Silvia Remus\inst{\ref{inst:usm}}
    \and
    Klaus Dolag\inst{\ref{inst:usm},\ref{inst:mpa}}
    \and
    Benjamin A. Seidel\inst{\ref{inst:usm}}
}
\authorrunning{L. M. Valenzuela et al.}

\institute{
    Universitäts-Sternwarte, Fakultät für Physik, Ludwig-Maximilians-Universität München, Scheinerstr. 1, 81679 München, Germany\label{inst:usm}\\
    \email{lval@usm.lmu.de}
    \and
    Max-Planck-Institut für Astrophysik, Karl-Scharzschild-Str. 1, 85748 Garching, Germany\label{inst:mpa}\\
}

\date{Received XX Month, 20XX / Accepted XX Month, 20XX}

\abstract
{
Despite being one of the most fundamental properties of galaxies that dictates the form of the potential, the 3D shapes are intrinsically difficult to determine from observations. The improving quality of triaxial modeling methods in recent years has made it possible to measure these shapes more accurately.
}
{
This study provides a comprehensive understanding of the stellar and dark matter (DM) shapes of galaxies and the connection between them. As these shapes are the result of the formation history of a galaxy, we investigate which galaxy properties they are correlated with, which will be especially useful for interpreting the results from dynamical modeling.
}
{Using the hydrodynamical cosmological simulation \emph{Magneticum Pathfinder Box4 (uhr)}, we computed the stellar and DM intrinsic shapes of 690~simulated galaxies with stellar masses above \SI{2e10}{\Msun} at three different radii with an iterative unweighted method. We also determined their morphologies,
their projected morphological and kinematic parameters, and their fractions of in-situ formed stars.}
{
The DM follows the stellar component in shape and orientation at three half-mass radii, indicating that DM is heavily influenced by the baryonic potential in the inner parts of the halo.
The outer DM halo is independent of the inner properties such as the DM shape or galaxy morphology, however, and is more closely related to the large-scale anisotropy of the gas inflow.
Overall, DM halo shapes are prolate, consistent with previous literature.
The stellar shapes of galaxies are correlated with their morphology, with early-type galaxies featuring more spherical and prolate shapes than late-type galaxies out to \SI{3}{\Rhalf}.
Galaxies with more rotational support are flatter, and the stellar shapes are connected to the mass distribution, though not to the mass itself. In particular, more extended early-type galaxies have larger triaxialities at a given mass.
Finally, the shapes can be used to better constrain the in-situ fraction of stars when combined with the stellar mass.
}
{The relations between shape, mass distribution, and in-situ formed star fraction of galaxies show that the shapes depend on the details of the accretion history through which the galaxies are formed. The similarities between DM and stellar shapes in the inner regions of galaxy halos signal the importance of baryonic matter for the behavior of DM in galaxies and will be of use for improving the underlying assumptions of dynamical models for galaxies in the future.
However, at large radii the shapes of the DM are completely decoupled from the central galaxy, and their shapes and spin are coupled more to the large scale inflow than to the galaxy in the center.
}


\maketitle
%

\section{Introduction}
\label{sec:introduction}

The large variety of observed galaxies has led to various approaches of classifying galaxies. As the first such classification scheme, the Hubble sequence \citep{hubble26,hubble36} distinguishes elliptical or early-type galaxies (ETGs) from spiral or late-type galaxies (LTGs). The ETG class is further subdivided by using the projected ellipticity. Even with some additions and modifications \citep[e.g.,][]{de_vaucouleurs59,liller66,kormendy&bender96}, this general distinction of galaxy types has remained the same over the years.
This type of classification based on visual features was crucial to advance the understanding of galaxies and their formation through the tight coupling of the morphology and the formation history of galaxies.

Through the advancement of integral field spectroscopy (IFS), especially ETGs are nowadays oftentimes classified by their kinematics and projected shape, for example by separating them into fast and slow rotators \citep[e.g.,][]{cappellari+07:sauronX,emsellem+07:sauronIX,emsellem+11:atlas3dIII}. A significant advantage compared to the subdivision by ellipticity is that the overall structure of the line-of-sight (LOS) kinematics is less sensitive to the inclination of the galaxy compared to the projected ellipticity. The ellipticity and kinematics have been correlated in previous research, for example through relating the ellipticity and rotational support from the LOS velocity distribution with the global anisotropy \citep[e.g.,][]{illingworth77,binney78,binney05,emsellem+07:sauronIX,emsellem+11:atlas3dIII}.

Technical advancements have also led to improved measurements of the low surface brightness regions in the outskirts of galaxies. Since the relaxation timescales are larger in those areas dominated by the stellar halo, the details of the stellar distribution at large radii are expected to show imprints of the formation history and evolution. Many studies have targeted the outskirts of galaxies -- initially only through the use of tracer populations, such as globular clusters and planetary nebulae \citep[e.g.,][]{hui+95,arnaboldi+96,peng+04:GCsII,peng+04:PNe,arnaboldi+12,pota+13,versic+24}, and more recently also using stellar properties out to larger radii \citep[e.g., through the SLUGGS survey,][]{brodie+14:sluggs,arnold+14,foster+16,dolfi+21}.
Furthermore, a number of deep imaging surveys have also enabled the study of low surface brightness stellar features in the halo, such as tidal tails, streams, and shells (e.g., \citealp{duc+15:atlas3dXXIX,bilek+20,martinez_delgado+23,sola+22,valenzuela&remus24,rutherford+24}).

While the first morphological classification schemes were all based on the 2D projected measurements obtained from observations, the 3D spatial distribution of matter is the actual underlying physical property.
In dynamical modeling of galaxies, the shapes of the DM and stellar components are central for modeling the potential, for instance in Schwarzschild orbital modeling \citep{schwarzschild79:schwarzschild_modeling,schwarzschild82:schwarzschild_modeling} or Jeans Anisotropic Multi-Gaussian Expansion (JAM) dynamical modeling \citep{cappellari08:jam}.
The 3D \enquote{shape} of the underlying stellar potential has oftentimes been approximated by spherical \citep[e.g.,][]{richstone&tremaine85,rix+97} or axisymmetric spheroidal models \citep[e.g.,][]{van_der_marel+98,gebhardt+03,krajnovic+05,walker+09,li+16,versic+24}, but has moved towards triaxial ellipsoidal descriptions \citep[e.g.,][]{van_den_bosch+08,van_de_ven+08,jethwa+20:dynamite,vasiliev&valluri20,pilawa+22:massiveXVII}. However, the DM component itself is mostly still described by spherically symmetric models, such as through the NFW profile (\citealp{navarro+96:nfw}; e.g., \citealp{poci+21,poci&smith22,santucci+22}), generalized NFW profiles \citep[e.g.,][]{bellstedt+18}, or alternative spherical parametrizations \citep[e.g.,][]{derkenne+21,pilawa+22:massiveXVII,quenneville+22}. Only recently, some models have allowed for triaxial DM potentials \citep[e.g.,][]{neureiter+21:smart,de_nicola+22}.

Observations and simulations have long indicated that neither galaxies nor their halos are spherically symmetric or axisymmetric, but triaxial. Early observations of the projected kinematics and the isophote shapes have hinted at triaxial shapes of the stellar component \citep[e.g.,][]{illingworth77,bender88,wagner+88,franx+91,vincent&ryden05}. N-body simulations were also able to produce triaxial stellar shapes \citep[e.g.,][]{aarseth&binney78,hohl&zang79,barnes92}, as well as triaxial DM halos \citep[e.g.,][]{jing&suto02}. Cosmological simulations have since shown that DM halos are actually expected to be triaxial \citep[e.g.,][]{springel+04,bailin&steinmetz05,allgood+06,bett+07,schneider+12,tenneti+14,bonamigo+15,hellwing+21}, and that the stellar component can have triaxial shapes as well \citep[e.g.,][]{tenneti+14,pulsoni+20,de_nicola+22:shapes}. Such studies have found that more massive galaxies have more triaxial shapes, and that DM is generally more spherical than the stellar component.
Other studies have also analyzed the evolution of DM shapes, where \citet{vera_ciro+11} found that the DM halo of Milky Way-mass systems becomes more triaxial and oblate over time. Furthermore, the alignment of galaxies and their halos and of halos with the surrounding cosmic web have been analyzed to find how the large-scale structures influence the smaller scales \citep[e.g.,][]{bett12,forero_romero+14,hellwing+21,xu+23,xu+23:tng}.

Clearly, it is essential to have a good understanding of the 3D shapes of the stellar and DM components in galaxies. First, dynamical modeling relies on accurate descriptions of the underlying potential, for which it has been shown that spherical symmetry is inaccurate. Second, connecting the physical 3D shape to galaxy kinematics is expected to carry more information on the formation history than the projected shapes alone, particularly for the low surface brightness galaxy outskirts. Last, connections between the not directly measurable DM and the stellar shape hold the potential of new insights into the nature of DM and its role in galaxy physics.

Shapes have been measured and studied in a number of hydrodynamical cosmological simulations.
The shapes and alignments of the stellar and DM components have been studied for the galaxy halos in the MassiveBlack-II simulation by \citet{tenneti+14} and for the galaxies and halos in the cosmo-OWLS and EAGLE simulations by \citet{velliscig+15} and \citet{petit+23}. \Citet{tenneti+14} found that the total stellar and DM halos are overall well aligned, and \citet{velliscig+15} concluded that the galaxies align well with the local total matter orientation out to \SI{1}{\Rvir}, in particular for LTGs, and that the inner galaxy is oftentimes misaligned with the outer halo orientation for systems below galaxy group masses. However, both studies determined the shapes using a non-iterative method, for which the physical meaning of the resulting \enquote{shapes} is unclear \citep{zemp+11}. \Citet{petit+23} focused on the halo shapes at $z=0.5$ and found better overall alignment between the DM and gas halos than between the DM and stellar halos.
For a fixed radius of \SI{30}{\kilo\parsec}, \citet{thob+19} studied the relations between the galaxy shapes and the kinematics in EAGLE, finding a relation of more rotationally-supported galaxies being flatter, but with a significant scatter of the shape for galaxies with less rotational support.
Studies have also shown that the inclusion of baryons in simulations leads to rounder DM halos, in particular in the inner halo regions \citep[e.g.,][]{bryan+13,nunez_castineyra+23}.

For the IllustrisTNG simulations, \citet{li+18:prolate} investigated prolate galaxies in TNG100, finding that prolate galaxies are massive ($M_* \gtrsim \SI{3e11}{\Msun})$ and have had a massive dry merger in their past. For TNG50 and TNG100, \citet{pulsoni+20,pulsoni+21} related the stellar shapes of ETG halos and their radial profiles out at large radii to the rotational support and to the classification of being a fast or slow rotator, with the aim of obtaining clues of the galaxy's formation history. Fast and slow rotators were found to behave similarly in the outskirts, while fast rotators tend to be more oblate in the center.
\Citet{emami+21} analyzed twists and stretched shapes of Milky Way-like galaxies in TNG50, finding that the inner DM and stellar shape profiles align well, indicative of how the DM responds to the baryonic component.
Finally, \citet{zhang+22} studied the shapes of quiescent galaxies in TNG50, finding more spheroidal shapes in low-mass galaxies and at lower redshifts. They also concluded that shape transformations occur through violent relaxation caused by mergers because of the high ex-situ stellar mass fractions seen in massive galaxies, which have more spheroidal shapes than lower-mass galaxies.

The mentioned studies have mostly analyzed the shapes of galaxy halos or only of a specific subset of galaxies, for which they have focused on the orientations in space and on certain kinematic properties.
The aim of this paper is to draw a more complete picture of the stellar and DM 3D shapes of galaxies and to connect them with further fundamental properties.
For this, we use the \emph{Magneticum Pathfinder} simulation suite to particularly study the behavior within a few half-mass radii of the galaxies, which is of high relevance for dynamical modeling, and in what way these inner regions are correlated to the halo properties.
The statistical study of the stellar and DM shapes in the inner and outer regions of galaxies and their relation to each other will further provide valuable insights useful for dynamical modeling, especially since deep imaging of the outskirts has become available for an increasing number of galaxies.
In \cref{sec:sim_methods}, we first introduce the simulation and the galaxy sample used in this work as well as the definitions and methods needed for the computed galaxy properties.
A statistical overview of the stellar and DM shapes is presented in \cref{sec:shape_properties}, after which they are connected with other galaxy and halo properties in \cref{sec:global_properties}. Finally, we analyze the shape alignments and the relation to galaxy dynamics in \cref{sec:alignment} and conclude in \cref{sec:summary_conclusion}.

\section{Simulation and methods}
\label{sec:sim_methods}

In the following, we first introduce the simulation we used for this study and describe how the galaxy sample was selected. Then, we present the detailed methods used to quantify the various galaxy properties that we investigated.

\subsection{The Magneticum Pathfinder simulations}
\label{sec:magneticum}

The galaxies studied in this work are taken from the \emph{Magneticum Pathfinder}\footnote{\url{www.magneticum.org}} simulation suite, a set of hydrodynamical cosmological simulations performed with \gadget{3}, an extended version of \gadget{2} \citep{springel05:gadget2}. It includes updates to the smoothed particle hydrodynamics (SPH) formulation \citep{dolag+04,dolag+05,donnert+13}, the active galactic nuclei (AGN) feedback \citep{fabjan+10,hirschmann+14}, and the star formation and metal enrichment \citep{tornatore+04,tornatore+07,wiersma+09}, for example. The simulations were performed using a standard \LCDM{} cosmology corresponding to the seven-year results of \wmap{} \citep{komatsu+11:wmap7}: $h = 0.704$, $\Omega_\Lambda = 0.728$, $\Omega_m = 0.272$, $\Omega_b = 0.0451$, $\sigma_8 = 0.809$, and $n_s = 0.963$.
The simulation used for this work is \MagneticumBox{4}{uhr, \emph{ultra-high resolution}}, which has a side length of $\SI{48}{\mega\parsec\per\h} \approx \SI{68}{\mega\parsec}$ and contains DM particles of mass $m_\mathrm{DM} = \SI{3.6e7}{\Msun\per\h}$ and gas particles of initial mass $m_\mathrm{gas} = \SI{7.3e6}{\Msun\per\h}$, which form stellar particles with average mass $m_* = \SI{1.3e6}{\Msun\per\h}$ (a gas particle can form up to four stellar particles). The same softening length is used for DM and gas particles, $\epsilon_\mathrm{DM} = \epsilon_\mathrm{gas} \approx \SI{2}{\kilo\parsec}$, whereas a smaller value is used for stellar particles, $\epsilon_* = \SI{1}{\kilo\parsec}$. The simulation is initialized with $2 \times 576^3$ particles. The size and resolution of the simulation provide a good balance between ensuring a sufficiently large galaxy sample while resolving enough details of galaxies down to a stellar mass of almost \SI{e10}{\Msun}.
Galaxies are identified through a version of \subfind{} adapted for baryonic matter \citep{springel+01:subfind,dolag+09:subfind}.

Previous studies have shown that the galaxies from \MagneticumBox{4}{uhr} successfully reproduce observational properties. These include their angular momenta \citep{teklu+15}, kinematics \citep{schulze+18,van_de_sande+19,schulze+20} and dynamics \citep{remus+17,teklu+17,harris+20,remus&forbes22} at different redshifts, structures in the outskirts \citep{foerster+22,valenzuela&remus24}, and their in-situ component fractions \citep{remus&forbes22}.

\begin{figure}
    \centering
    \includegraphics[width=\columnwidth]{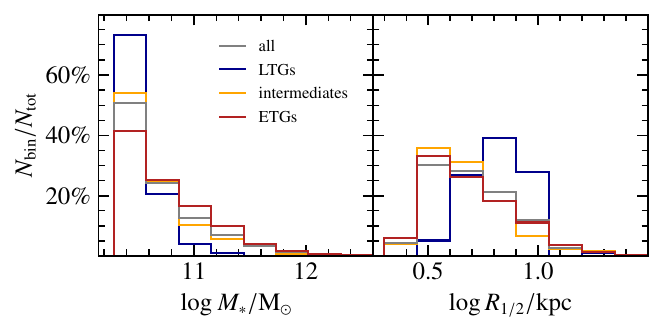}
    \caption{Mass and size distributions of the entire Magneticum galaxy sample as well as only for the ETGs, intermediate-type galaxies, and LTGs, using the morphological classification introduced in \cref{sec:morphology}.}
    \label{fig:mass_size_histograms}
\end{figure}

For this study, we selected the galaxy sample from the subhalos identified by \subfind{} in the last snapshot of the simulation ($z=0.06$) based on the stellar mass criterion of $M_* \geq \SI{2e10}{\Msun}$ and the stellar half-mass radius criterion $\Rhalf \geq \SI{2}{\kilo\parsec}$ (twice the softening length to ensure an adequate spatial resolution), which corresponds to the same base sample as used by \citet{schulze+18}. We further only consider the main subhalos (i.e., they are not satellite galaxies) and require the total DM mass to also satisfy $M_\mathrm{DM} \geq \SI{e11}{\Msun}$ to ensure a sufficient DM resolution, resulting in a sample of 690~galaxies.
We show the mass and size distributions of the galaxies in \cref{fig:mass_size_histograms} to characterize the sample, as well as the distributions for the individual morphological types, divided into ETGs, intermediate-type galaxies, and LTGs.

Due to the significant difference in stellar and DM particle mass, the stellar component is collisionally heated through the more massive DM particles, leading to low-mass galaxies being puffed up \citep[e.g.,][]{ludlow+21,ludlow+23,wilkinson+23}. Based on the work by \citet{ludlow+23}, this should only affect galaxies with total masses of around $M_\mathrm{tot} \lesssim \SI{6e11}{\Msun}$--\SI{1e12}{\Msun}, which corresponds to the lower-mass end of our sample. For this reason, the low-mass galaxies are expected to be less flattened than they would actually be. As this is a systematic effect, however, the relations studied in this paper still hold and may even be slightly more prominent in reality.

\subsection{Galaxy shapes}
\label{sec:shapes}

To determine the shapes of galaxies, a large variety of methods have been used in the literature (see \citealp{zemp+11} for an overview), which generally are based on determining a best-fitting ellipsoid by using the second moment of the mass distribution tensor, $\tens{M}$:
\begin{equation}
    \label{eq:mass_tensor}
    M_{ij} = \sum_{k} \tilde{m}_k w(\vec{r}_k) r_{k,i} r_{k,j},
\end{equation}
where the sum runs over all considered particles $k$ within an initial sphere of a given radius (which may iteratively be deformed as discussed below), $\tilde{m}_k$ is the particle mass, $\vec{r}_k$ is the particle position relative to the host galaxy's origin (with $r_{k,i}$ being the $i$-component of the vector; the method for identifying the galaxy's origin is described in \cref{sec:com_cov}), and $w(\vec{r}) = 1$ is a weighting function. The corresponding ellipsoid's axis ratios $q = b/a$, $s = c/a$, and $p = c/b$ (with the half-axis lengths $a \geq b \geq c$) are given by the square roots of the eigenvalue ratios of $\tens{M}$: $q = \sqrt{\lambda_2 / \lambda_1}$, $s = \sqrt{\lambda_3 / \lambda_1}$, and $p = \sqrt{\lambda_3 / \lambda_2}$, with the eigenvalues $\lambda_1 \geq \lambda_2 \geq \lambda_3$.

For variations of the tensor $\tens{M}$, $\tilde{m}_k$ can be set to 1 or another constant value for a non-mass-weighted determination of the tensor, and the weighting function $w(\vec{r})$ can be used to weight particles according to their distances from the galaxy's center, for example with $w(\vec{r}) = r^{-2}$ or $w(\vec{r}) = r_\mathrm{ell}^{-2}$. The ellipsoidal radius is given by
\begin{equation}
    \label{eq:rell}
    r_\mathrm{ell}^2 = x_\mathrm{ell}^2 + \frac{y_\mathrm{ell}^2}{q^2} + \frac{z_\mathrm{ell}^2}{s^2},
\end{equation}
where $x_\mathrm{ell}$, $y_\mathrm{ell}$, and $z_\mathrm{ell}$ are the coordinates in the eigenvector system of the ellipsoid fitted to the galaxy.

The best-fitting ellipsoid can be determined iteratively by computing $\tens{M}$ through summing only over particles located within the ellipsoid determined in the previous iteration step. For example, to determine the ellipsoidal shape at a particular distance, $\tens{M}$ is computed from all particles within a sphere with a radius of that distance. For the determined axis ratios $q$, $s$, and $p$, and in the ellipsoid's eigenvector space, only the particles within an ellipsoid given by these parameters are considered in the next step. For this, there are the options to either keep the ellipsoid's major axis length, $a$, or the volume of the ellipsoid constant.

An alternative possibility of determining the best-fitting shape is to only consider particles within a homoeoid (an ellipsoidal shell, as opposed to an ellipsoid) at a given radius to obtain $\tens{M}$. Because of the limited resolution of the simulations and the large mass range of galaxies included in the sample, we followed the conclusion from \citet{zemp+11} to use the full ellipsoidal volume methods instead of homoeoids because of not being able to guarantee a few thousand particles per ellipsoidal shell. The galaxies used in this work also have sufficiently steep gradients in the regions that the shapes are measured, such that the resulting shapes are physically meaningful \citep{fischer&valenzuela23}.

For this study, we present a test of the robustness of these options in \cref{app:shape_methods}, where we considered the unweighted, reduced, and reduced ellipsoidal methods applied to ellipsoids, as well as their iterative (keeping the volume or major axis length constant) and non-iterative variations. We found that the reduced and non-iterative methods have a significant bias towards more spherical shapes, the reduced ellipsoidal methods are a good indicator for the average shape within the boundary, and the constant-axis length methods make a consistent comparison between equal-mass galaxies more difficult than the constant-volume methods. For these reasons, we concluded that for the selected galaxy sample, the best approach to determine the local shape at a given distance is the iterative unweighted method at constant volume for particles within an ellipsoid:
\begin{equation}
    M_{ij} = \sum_{k} m_k r_{k,i} r_{k,j}.
\end{equation}
This method ensures that particles at the center do not dominate the sum by being unweighted and that the determined shapes are not biased towards being spherical through the iterative computation.
We determined the shapes by only considering particles attributed to the respective subhalo from \subfind{} to prevent substructure from biasing its shape.

In addition to the axis ratios, an additional measure for the shape is the triaxiality, $T$. The triaxiality is defined as
\begin{equation}
    T = \frac{1 - q^2}{1 - s^2}
\end{equation}
and has values between 0 and 1. Values of $0 \leq T < 1/3$ indicate \emph{oblate} shapes, $1/3 \leq T < 2/3$ \emph{triaxial} shapes, and $2/3 \leq T \leq 1$ \emph{prolate} shapes. For nearly spherical shapes with $s$ being close to 1, the exact value of the triaxiality loses in meaning, however. This is a result of the fact that minor changes to one of the axis ratios leads to large changes of the triaxiality in that case.

\subsection{Center of particles and velocity}
\label{sec:com_cov}

As the choice of center of a galaxy has a large effect on fitting an ellipsoidal shape centered around it and on the resulting angular momenta, it is necessary to determine a reasonable galaxy center.
First, we apply the shrinking-sphere method from \citet{power+03:shrinking_sphere} to the stellar component: the sphere is initialized with a radius of $3\,R_{1/2,\mathrm{init}}$ (where $R_{1/2,\mathrm{init}}$ is the initial stellar half-mass radius determined around the point of the deepest potential; see \cref{sec:stellar_properties}) and is shrunk iteratively by \SI{2.5}{\percent} until the number of stellar particles within the sphere reaches \num{1000} particles, \SI{1}{\percent} of the particles within the initial sphere, or the sphere's radius reaches $0.5\,R_{1/2,\mathrm{init}}$. In each iteration, the center of the sphere is shifted to the center of mass of the contained particles. The actual stellar half-mass radius is then computed around the final center. This origin is used for both the stellar and DM components, but is only determined from the stellar component because the potential in the inner regions is dominated by it and this is the center that one can obtain directly from observations of galaxies.
This is further supported by the small mean relative offset between our determined galaxy center and the galaxy position obtained from the subhalo finder of \SI{0.05 \pm 0.05}{\Rhalf}.

As we investigated the kinematic properties of the galaxies as well, we also required a robust method for obtaining a velocity frame of reference per galaxy.
Substructure that is actually passing by but is erroneously attributed to the subhalo by \subfind{} can potentially influence a simple mass-weighted mean velocity strongly. For this reason, to find a velocity frame of reference for the galaxy (which can move freely around in the cosmological box) that is not skewed by such effects, we compute the mass-weighted mean velocity from the stellar particles except those with the \SI{10}{\percent} highest absolute velocities relative to the median velocity of the considered particles.
Additionally, only particles within four stellar half-mass radii are taken into account.

\subsection{Other galaxy properties}
\label{sec:properties}

We also used a variety of quantities that describe different aspects of a given galaxy in this work, including intrinsic 3D properties as well as projected and large-scale properties. These are introduced in the following.

\subsubsection{Intrinsic stellar properties}
\label{sec:stellar_properties}

The stellar mass, $M_*$, was determined from all stellar particles attributed to the respective subhalo (i.e., excluding substructure) within \SI{0.1}{\Rvir} (where $\Rvir$ is the virial radius as determined by \subfind{}), following the approach by \citet{teklu+15} and \citet{schulze+18,schulze+20}. The stellar half-mass radius was then obtained from these particles by determining the radius within which half of the mass of those particles resides. These properties have been shown to agree well with observations \citep[e.g.,][]{schulze+18,harris+20}.

For the in-situ fraction, $f_\text{in-situ}$, the fraction of stellar particles that were formed within a given main galaxy (independent of where the gas came from), we followed the procedure of \citet{remus&forbes22} by tagging particles based on their formation site.
Lastly, we determined the mean stellar age, $\langle t_{\mathrm{age},*} \rangle$, from the stellar particles within one half-mass radius.

\subsubsection{Projected stellar properties}
\label{sec:projected_properties}

\begin{figure}
    \centering
    \includegraphics[width=0.9\columnwidth]{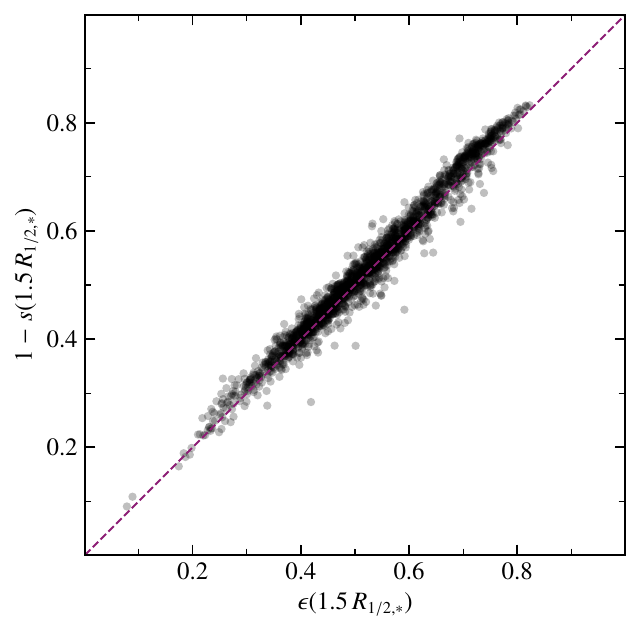}
    \caption{Edge-on ellipticity at \num{1.5}~half-mass radii of the galaxy sample compared to the axis ratio $s$ from the 3D ellipsoid. The edge-on view for the ellipticity was determined from the best-fitting ellipsoid at \num{1.5}~half-mass radii.}
    \label{fig:shape_2d_3d}
\end{figure}

When comparing simulated with observed galaxies, it is necessary to resort to properties determined from a projected view of a given galaxy. For galaxy shapes, this means that the full 3D information contained in $q$ and $s$ is not readily available for observed galaxies. However, a 2D ellipse can be fitted to the projected particle data in the same way as ellipsoids to the 3D particle positions (as presented in \cref{sec:shapes}).
A parameter that is commonly used to describe this ellipse is the \emph{ellipticity}, which is defined as
\begin{equation}
    \epsilon = 1 - \frac{b}{a},
\end{equation}
where $a \geq b$ are the half-axis lengths.
For a galaxy viewed edge-on, this parameter corresponds to $\epsilon \approx 1 - s$, and for a face-on galaxy to $\epsilon \approx 1 - q$, where $q$ and $s$ are the ellipsoidal axis ratios introduced in \cref{sec:shapes}. In \cref{fig:shape_2d_3d} we show this nearly 1:1-relation for the edge-on case, where we determined $\epsilon$ with the iterative unweighted method keeping the area of the deforming ellipse constant.
This means that the 3D shape parameters can be used for direct comparisons with observations when there are estimates available for the inclination of observed galaxies.

Additionally, methods have been developed to quantify the projected kinematics of galaxies. A 2D proxy for the angular momentum is the \lambdaR{}-value, first introduced by \citet{emsellem+07:sauronIX} as a part of the \sauron{} project \citep{bacon+01:sauronI}.
The parameter quantifies the rotational support of a galaxy and by how much its kinematics are dominated by dispersion. For observations, \lambdaR{} is defined as
\begin{equation}
    \label{eq:lambda_r}
    \lambda_R = \frac{\langle R |V| \rangle}{\big\langle R \sqrt{V^2 + \sigma^2} \big\rangle},
\end{equation}
where $R$ is the projected distance to the galaxy's center, $V$ and $\sigma$ are the 1D line-of-sight velocity and velocity dispersion, and the angle brackets $\langle \cdot \rangle$ represent the luminosity-weighted average.

The \lambdaR{}-parameter can be determined for simulations assuming a constant mass-to-light ratio (following e.g., \citealp{jesseit+09,naab+14:atlas3dXXV,schulze+18}):
\begin{equation}
    \label{eq:lambda_r_sim}
    \lambda_R = \frac{\sum_i M_i R_i |V_i|}{\sum_i M_i R_i \sqrt{V_i^2 + \sigma_i^2}},
\end{equation}
where the sums run over 2D bins obtained through the centroidal Voronoi tessellation binning method\footnote{The centroidal Voronoi tessellation binning method ensures roughly circular bins of pixels, an approximately constant signal-to-noise ratio across all bins, and maximizes the spatial resolution.} \citep{cappellari&copin03:cvt}, and the quantities $M_i$, $R_i$, $V_i$, and $\sigma_i$ are the total mass, distance from the galaxy's center, mean velocity, and mean velocity dispersion of the $i$th Voronoi bin, respectively.
For further details on the properties of \lambdaR{} in the Magneticum simulations, see \citet{schulze+18,schulze+20}.
For this work, we calculated \lambdaR{} following this method and using \cref{eq:lambda_r_sim}.

When viewed from an edge-on perspective, galaxies dominated by rotation have values of $\lambdaR \sim 1$, whereas entirely dispersion-dominated galaxies have values of $\lambdaR \sim 0$. Assuming that the rotation is orthogonal to the intermediate principal axis of the galaxy (i.e., around the minor or major axis), \lambdaR{} is expected to be maximal in the edge-on projection and then represents the true amount of rotational support.

Finally, we group the galaxies into four different groups based on the kinematic classifications of the Magneticum galaxy sample by \citet{schulze+18}, who classified the galaxies according to their edge-on projected line-of-sight velocity maps within one half-mass radius. As \citet{valenzuela&remus24}, we distinguish between four groups:
\begin{itemize}
    \item Regular rotators: ordered rotation around the minor axis,
    \item Non-rotators: no ordered rotation visible,
    \item Kinematically distinct cores: ordered rotation in the very inner regions with non-rotational or differently oriented rotation in the outskirts (this class consists of the distinct core and kinematically distinct core groups from \citet{schulze+18}),
    \item Prolate rotators: ordered rotation around the major axis.
\end{itemize}
More details on the classification and the individual properties of these groups of galaxies are presented by \citet{schulze+18}.

\subsubsection{Morphology}
\label{sec:morphology}

The morphology is one of the most fundamental properties observed for galaxies.
Based on the findings that LTGs and ETGs follow parallel linear relations in the logarithmic $M_*$-$j_*$-plane (where $j_*$ is the specific stellar angular momentum) depending on the disk-to-bulge ratios \citep[e.g.,][]{fall83,fall&romanowsky13}, \citet{teklu+15} showed that the morphological type of a galaxy can be quantified through the $y$-intercept of the relation, the \bvalue{}:
\begin{equation}
    \label{eq:bvalue}
    b = \log\bigg( \frac{j_*}{\si{\kilo\parsec\kilo\meter\per\second}} \bigg) - \frac{2}{3} \log\bigg( \frac{M_*}{\si{\Msun}} \bigg).
\end{equation}
Following \citet{teklu+17} and \citet{schulze+20}, we define the morphological class of the simulated galaxies as: 
\begin{itemize}
    \item LTG: $b \geq -4.35$,
    \item Intermediate: $-4.73 < b < -4.35$,
    \item ETG: $b \leq -4.73$.
\end{itemize}

\subsubsection{Gas inflow field asphericity}
\label{sec:asphericity}

As the scale of a galaxy's halo is much larger than that of the inner galaxy itself, it is useful to quantify some of the large-scale properties beyond only the size and total mass of the halo. Therefore, we want to quantify how isotropic or anisotropic the infall of matter from the environment is, which could directly influence the shape of the halo.
For this work, we used the asphericity of the gas inflow field on the spherical surface at one virial radius.

First, the density and velocity fields of the inflowing gas particles are determined (i.e., gas particles with negative radial velocities) through a spherical surface with a radius of \SI{1}{\Rvir} using a modified version of the post-processing program SMAC \citep{dolag+05:smac}.
Using an SPH kernel interpolation, this results in two HEALPix \citep{gorski+05:healpix} maps (density and inflow velocity) per galaxy that are subsequently multiplied pixel-by-pixel and additionally multiplied with the pixel surface area to produce the instantaneous inflow field through the spherical surface. This procedure is described in further detail by Seidel et al.\ (in prep.).

Using this inflow field, the \emph{asphericity} $\zeta$ of the inflow is then quantified by first performing a spherical harmonics decomposition on the map and then calculating the ratio of the summed power in the higher-order moments ($0<l<10$) to the monopole power ($l=0$). In general, more aspherically distributed fields have higher such ratios, that is higher values of $\zeta$. Similar procedures have been used in the past both in 3D for studying how aspherical the inflows are in isolated galaxy cluster settings \citep{valles_perez+20}, for determining the preferred infall directions for satellite galaxies in high-resolution zoom-in simulations of the local group \citep{libeskind+11}, and in 2D projection to study possible correlations of the asphericity of the gas distribution in cluster outskirts with measures of the dynamical state \citep{gouin+22}.

\section{Shape properties}
\label{sec:shape_properties}

We applied the selected shape determination method (iterative unweighted at constant volume for particles within an ellipsoid; see \cref{sec:shapes} and \cref{app:shape_methods}) to the simulated galaxy sample, obtaining the axis ratios $q$ and $s$ as well as the triaxiality $T$ for each galaxy. These were computed at different radii for the stellar and DM components, which is detailed in the following. The relations between the three shape parameters are presented and discussed as well.

\subsection{Stellar shapes}
\label{sec:stellar_shapes}

\begin{figure}
    \centering
    \includegraphics[width=0.9\columnwidth]{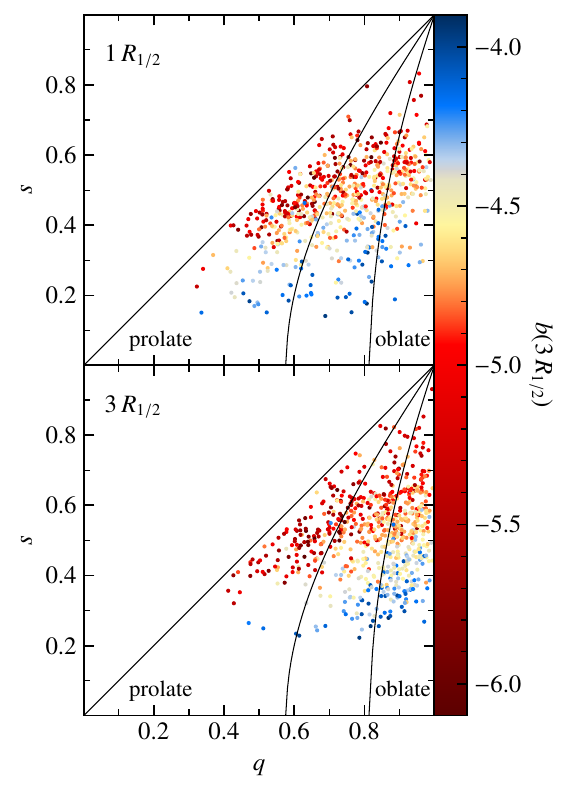}
    \caption{Axis ratios $q$ and $s$ of the stellar component of the galaxy sample at one and three stellar half-mass radii. The data points are colored by the \bvalue{}. The solid lines indicate the borders between prolate, triaxial, and oblate shapes, from left to right, respectively.}
    \label{fig:q_s_b_stellar}
\end{figure}

\begin{figure*}
    \centering
    \includegraphics{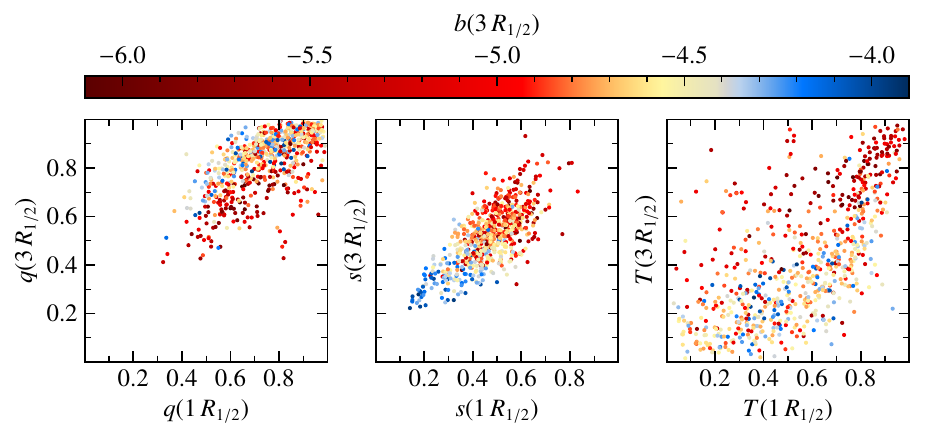}
    \caption{Stellar shape parameters $q$, $s$, and $T$ (from left to right, respectively) compared between the inner and outer regions measured at \SI{1}{\Rhalf} and \SI{3}{\Rhalf} for the simulated galaxy sample, colored by the \bvalue{}.}
    \label{fig:shapes_inner_outer_stellar}
\end{figure*}

Galaxy shapes differ depending on the radius at which they are measured. For the simulated galaxies, two reasonable radii for determining the shape are \SI{1}{\Rhalf} and \SI{3}{\Rhalf} for probing the inner and outer shapes of the stellar component.
\Cref{fig:q_s_b_stellar} shows the distribution of shape parameters for the simulated galaxy sample and their morphologies in the $q$-$s$-plane at the two considered radii. Since the triaxiality is obtained directly from $q$ and $s$, the location of a galaxy in the plane indicates the triaxiality, with galaxies on the left being prolate and those on the right being oblate.

A clear relation is immediately visible between the axis ratios and the morphology at both radii: LTGs have a low value of $s$, while ETGs have high values of $s$ and are thus more spherical. This trend is expected due to the nature of elliptical versus disk galaxies. There is also a slight correlation with the face-on axis ratio, $q$. Comparing the shapes at \SI{1}{\Rhalf} and \SI{3}{\Rhalf}, we found that galaxies tend to be triaxial to prolate in the inner regions and oblate in the outskirts (also see \cref{fig:shapes_inner_outer_stellar}).
\Citet{pulsoni+20} found a similar distribution of shape axis ratios at \SI{8}{\Rhalf} in IllustrisTNG as we do at \SI{3}{\Rhalf} (note that their $p$ and $q$ correspond to $q$ and $s$ as used in this work, respectively), which both reside in the stellar halo. IllustrisTNG appears to have more flattened stellar halos than Magneticum, but interestingly also features more near-spherical halos. The differences could be influenced by the different feedback models employed for the stars and for AGN, but could also be the result of the different radii at which the stellar halo shapes were measured. For the AGN feedback, Magneticum employs the feedback model from \citet{fabjan+10} with modifications to the black hole pinning by \citet{hirschmann+14}, where the feedback efficiency is set to be higher for low accretion rates to mimick an efficient radio mode in the more quiescent phases, thus resulting in two AGN feedback modes. IllustrisTNG also employs two modes of feedback as introduced by \citet{weinberger+17}, where the radio mode is implemented through kinetic winds, which couple to the momentum of the gas instead of to its thermal properties. 
This and smaller differences in the treatment of stellar and supernova winds lead to changes in the timing of feedback and thus results in entirely different time evolutions over cosmological time scales. For this reason, the shapes at $z=0$ cannot be expected to have the same properties, but at the same time it is nearly impossible to trace back what exactly led to these differences. Further targeted work will be needed to better understand how individual feedback parameters influence these shapes.

\Citet{pulsoni+20} found that slow rotators generally have constant or decreasing triaxiality profiles, whereas fast rotators tend to have constant or increasing triaxialities.
However, it should be noted that the radii of \SIlist{1;3}{\Rhalf} correspond to the innermost range of radii at which \citet{pulsoni+20} measured their shapes, as they primarily studied the outer halo regions. A detailed analysis of the radial shape profiles in Magneticum computed out to larger radii will be presented in a second paper (Valenzuela et al.\ in prep.). 
From Schwarzschild dynamical modeling of observed ETGs in the MaNGA survey \citep{bundy+15:manga}, \citet{jin+20} found slightly more spherical and less oblate systems at \SI{1}{\Reff} (the Petrosian half-light radius), which is likely the result of the orbit families of which the spheroidal bulge component is composed.

\begin{figure}
    \centering
    \includegraphics[width=0.76\columnwidth]{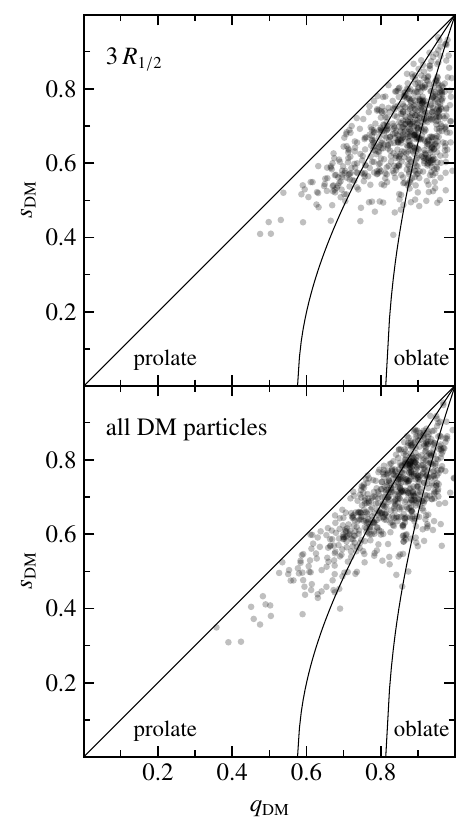}
    \caption{Axis ratios $q$ and $s$ of the DM component of the galaxy sample at three stellar half-mass radii and computed for all DM particles identified for the given galaxy by \subfind{}. The points are plotted semi-transparently to better show their distribution. The solid lines indicate the borders between prolate, triaxial, and oblate shapes.}
    \label{fig:q_s_dm}
\end{figure}

\begin{figure*}
    \centering
    \includegraphics[width=0.95\textwidth]{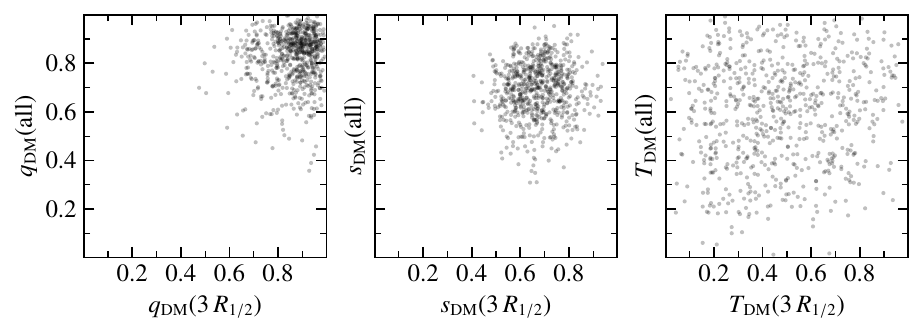}
    \caption{DM shape parameters $q$, $s$, and $T$ (from left to right, respectively) compared between the inner and outer halo regions for the simulated galaxy sample. The inner DM shape is determined at \SI{3}{\Rhalf} and the outer halo shape is computed from all DM particles belonging to the halo.}
    \label{fig:shapes_inner_outer_dm}
\end{figure*}

For LTGs, the higher triaxialities at \SI{1}{\Rhalf} can be interpreted as the galaxies being dominated by a bulge-like component in the inner regions, whereas the more oblate disk-like structure becomes apparent when probing the shapes further out, that is at \SI{3}{\Rhalf}. In contrast, the value of $s$ increases in the outskirts, which means that LTGs are less flat further out despite being more oblate. This can be explained by the presence of a more spherical stellar halo component. To obtain flatter shapes that resemble a disk for LTGs, it would be necessary to decompose the particles into multiple morphological components, which is beyond the scope of this work. Finally, for ETGs with especially low \bvalues{}, there is a clear preference for them being prolate at both radii.
\Cref{fig:shapes_inner_outer_stellar} shows that the shape parameters in the inner and outer regions of galaxies have similar values of $s$, whereas the face-on axis ratio $q$ has a larger scatter. There is also a tendency for the face-on projection of galaxies to be more circular in the outskirts, meaning that $q$ is larger further out, leading to the more oblate shapes at \SI{3}{\Rhalf}.

\subsection{Dark matter shapes}
\label{sec:dm_shapes}

\Cref{fig:q_s_dm} shows the DM shape parameters at three stellar half-mass radii and for the total DM component (there is generally not a sufficient number of DM particles within \SI{1}{\Rhalf} because of the larger DM softening length, especially for the smaller galaxies in the sample). The DM is overall much more spherical than the stellar component. This is especially true for the shapes at \SI{3}{\Rhalf}. With respect to the total DM shapes, most halos are still very spherical, but there is an increasing population of flattened and elongated systems, resulting in prolate DM halos. Their trends with the total mass and the large-scale structure are analyzed and discussed in \cref{sec:dm_global_properties}.
The total DM halo is generally also much less oblate than the inner DM component. In both cases, there is no correlation with the \bvalue{}, that is the morphology of the galaxy within. The findings of prolate total DM halo shapes are consistent with previous work done on DM halos in DM-only and hydrodynamical cosmological simulations \citep[e.g.,][]{allgood+06,bett+07,bonamigo+15,butsky+16,prada+19}.

For the inner parts of the DM halo, the axis ratios $q$ and $s$ are expected to overall be higher compared to DM-only simulations as a result of the influence of baryonic matter \citep[e.g.,][]{butsky+16,chua+19,prada+19,cataldi+21}. Our findings at \SI{3}{\Rhalf} agree well with the values obtained at different radii within the virial radius from such studies, with the majority of galaxies having values of $q \gtrsim 0.7$ and $0.6 \lesssim s \lesssim 0.8$.

When comparing the axis ratios between the DM shapes at \SI{3}{\Rhalf} and of the total DM halo, it becomes apparent that there is no correlation for either of the axis ratios or the triaxiality (\cref{fig:shapes_inner_outer_dm}). This means that the inner shapes of the DM are decoupled from those of the halo. Therefore, we find that no conclusions can be drawn for the DM in the inner parts of a galaxy based on the shape of the DM halo and vice versa.

\begin{figure*}
    \centering
    \includegraphics[width=0.95\textwidth]{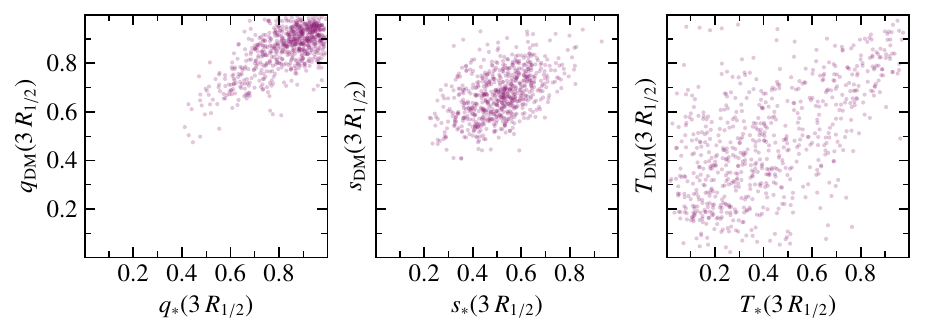}
    \caption{Stellar and DM shape parameters $q$, $s$, and $T$ compared with each other at a radius of \SI{3}{\Rhalf} from left to right, respectively.}
    \label{fig:shapes_stellar_dm}
\end{figure*}

\subsection{Stellar vs.\ dark matter shapes}
\label{sec:stellar_dm_shapes}

The important question now is to what extent the DM and its shape may be affected by the presence of baryonic matter. \Cref{fig:shapes_stellar_dm} shows that there is in fact a correlation between the stellar and DM shapes at \SI{3}{\Rhalf}, where $q$ shows the tightest relation. For the triaxiality, the tightest relation is found for prolate shapes, while the scatter is larger for triaxial and oblate shapes.
As is expected from \cref{sec:stellar_shapes,sec:dm_shapes}, the values of $q$ and $s$ are mostly higher for the DM than for the stellar component, leading to the more spherical shapes of DM.
These relations show that overall, the stellar and DM shapes are similar at \SI{3}{\Rhalf}, which agrees with previous studies that concluded that the DM and its shape are influenced by the baryonic potential \citep[e.g.,][]{chua+19,cataldi+21,emami+21}. In particular, our findings are consistent with the result from \citet{cataldi+21} that the DM is influenced by the baryons in the inner galaxy regions within \SI{20}{\percent} of the virial radius, whereas the triaxiality appears to be largely unaffected at one virial radius.

From \cref{fig:shapes_inner_outer_dm}, we found that the DM shapes in the inner and outer regions of halos are not related. We therefore conclude that despite being able to infer some properties of the DM shape in the inner halo regions through the stellar component, the decoupling of the outer shape means that the total DM halo shape properties remain unconstrained and are not deducible from observations of the inner galaxy.

\section{Correlating shapes with galaxy and halo properties}
\label{sec:global_properties}

Having studied the behavior and relations of the stellar and DM shapes at different radii, we next correlated the shape parameters $q$, $s$, and $T$ with the other galaxy properties introduced in \cref{sec:sim_methods}. Such relations have the potential to help constrain the underlying assumptions used for dynamical models of observed galaxies.

\begin{figure*}
    \centering
    \includegraphics{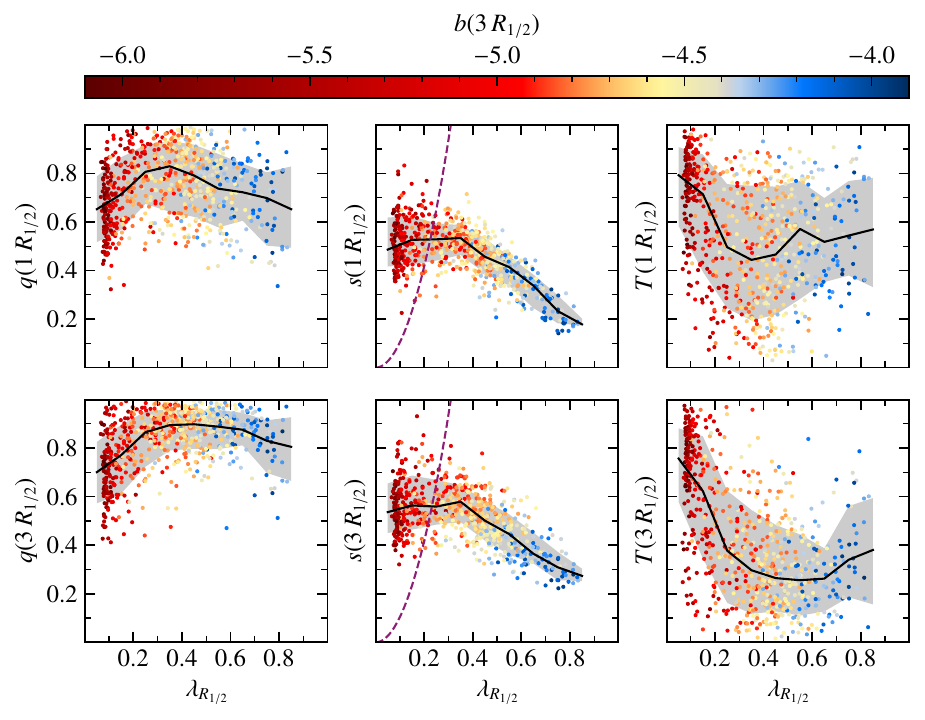}
    \caption{Relation between shape parameters at one (top row) and three (bottom row) stellar half-mass radii and \lambdaRhalf{}, colored by the \bvalue{}. The dashed blackberry lines in the middle column indicate the threshold between slow and fast rotators \citep{emsellem+11:atlas3dIII}, lying to the left and right of the line, respectively. The black lines indicate the median values in the respective \lambdaRhalf{} bins and the shaded regions the $1\sigma$ ranges (containing \SI{68}{\percent} of the galaxies above and below the median).}
    \label{fig:lambda_r_bvalue}
\end{figure*}

\subsection{Stellar component}
\label{sec:stellar_global_properties}

\begin{figure*}
    \centering
    \includegraphics{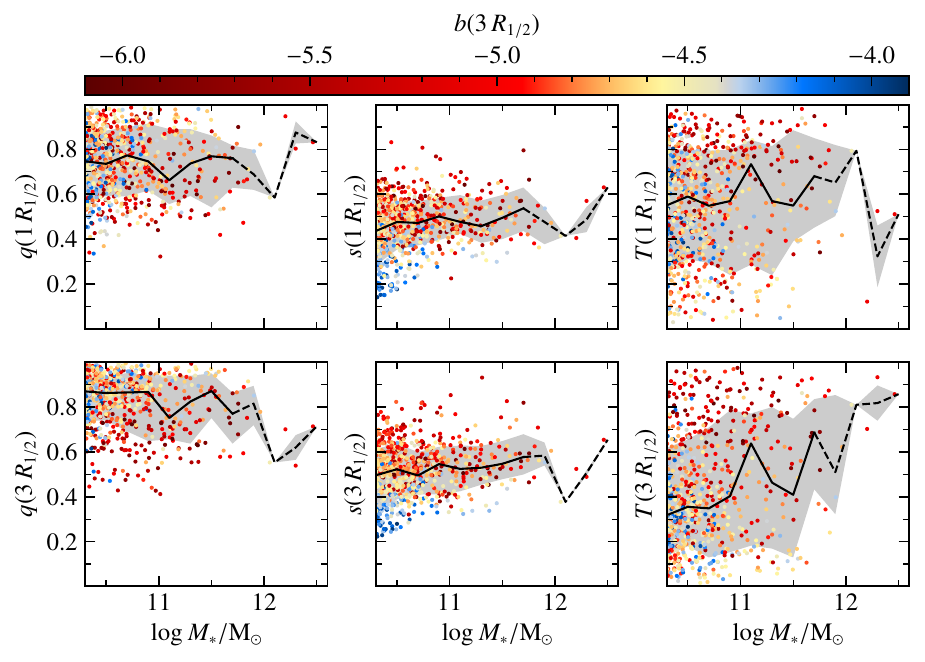}
    \caption{Relation between shape parameters at one (top row) and three (bottom row) stellar half-mass radii and the stellar mass, colored by the \bvalue{}. The black lines indicate the median values in the respective mass bins and the shaded regions the $1\sigma$ ranges (containing \SI{68}{\percent} of the galaxies above and below the median). The dashed part of the lines indicates where there may be effects caused by low number statistics.}
    \label{fig:mass_stellar}
\end{figure*}

Through the connection of galaxy shapes with the morphology (\cref{sec:stellar_shapes}) it is clear that the shapes are not entirely independent of other galaxy properties.
It has previously been shown in observations and simulations that the kinematic parameter \lambdaR{} is correlated with the projected ellipticity and the anisotropy, where the relation is especially tight for the edge-on projection \citep{illingworth77,binney78,binney05,emsellem+11:atlas3dIII,schulze+18}. Because of the direct correspondence of $s$ and the edge-on ellipticity (\cref{fig:shape_2d_3d}), this result can also be found for $s$. Since the morphology and the \lambdaR{}-value are tightly linked with ETGs having low \lambdaR{}-values and LTGs having high ones, any identified relations with the shape parameters are valid for both properties.

\Cref{fig:lambda_r_bvalue} shows how the shape parameters are related to the \lambdaR{}-value at one and three half-mass radii, where the middle column shows the equivalent to the \lambdaR{}-$\epsilon$-plane for edge-on galaxies.
For the face-on axis ratio, $q$, the galaxies with the most circular shapes are the intermediate galaxies, whereas ETGs and LTGs are slightly more elliptical at both tested radii. Because of the high values of $s$ for ETGs, this leads to especially prolate shapes for ETGs. In contrast, the flat LTGs with high values of $q$ are more oblate. The triaxiality of LTGs strongly depends on how large $q$ is: at \SI{1}{\Rhalf}, $q$ is around \num{0.7} and leads to comparably large triaxialities of \num{0.5}, while the very circular face-on shapes at \SI{3}{\Rhalf} lead to more oblate triaxialities of LTGs. The triaxial inner regions are potentially an imprint of the bulges or bars inside the disk centers, though the resolution is not sufficient to draw definite conclusions. The fact that intermediate galaxies feature the largest values of $q$ also leads to a dip in the triaxiality for those galaxies compared to the ETGs and LTGs.

In addition to the correlations with the morphology and rotational support, further properties of the stellar component that are linked to these also show correlations with the edge-on axis ratio, $s$. Through its relation in the \lambdaR{}-$\epsilon$-plane (see fig.\ 5 of \citealp{schulze+18}), this is the case for the anisotropy, where large anisotropies tend to occur for large values of $s$ and small anisotropies for small $s$. Since star-forming galaxies tend to be LTGs, we were also able to find that galaxies with high specific star-formation rates have small values of $s$. This is also true for galaxies with younger stellar ages in their central regions. In both cases, there is no correlation with $q$ or the triaxiality. As these results are a consequence of \cref{fig:q_s_b_stellar} and \cref{fig:lambda_r_bvalue}, they are shown in \cref{app:stellar_global_properties}.

For the stellar mass, there is no relation with the axis ratios, and only a slight trend with the triaxiality at \SI{3}{\Rhalf}, but with a large scatter (\cref{fig:mass_stellar}). The reason for the triaxiality increasing with stellar mass is that $q$ is minimally smaller for larger masses and is close to~1, leading to large changes of the triaxiality. This shows that the shape and the mass develop independently for galaxies and must be connected to different aspects of the formation history of galaxies. \Citet{pulsoni+20} found a similar upward trend of the triaxiality for both TNG50 and TNG100 towards higher masses, independent of the kinematics. The simulations all agree that fast rotating systems tend to have lower stellar masses than slow rotators and tend to be more oblate as well.
From Schwarzschild dynamical modeling of ETGs in the MaNGA survey \citep{bundy+15:manga}, \citet{jin+20} found a similar relation between the triaxiality and the stellar mass at \SI{1}{\Reff} as we do at \SI{3}{\Rhalf}, with median values rising between $T=0.3$ until $T=0.6$ for $\log M_* / \Msun = 10.5$--11.5. The difference in radius is likely related to the differences in how the spheroidal bulge component is observed and modeled compared to how it appears in the simulation. Similarly, the results are also consistent with those from \citet{thater+23} for \atlas{} \citep{cappellari+11:atlas3dI} galaxies. In contrast, \citet{santucci+22} found much more oblate galaxies when dynamically modeling galaxies from the SAMI (Sydney-Australian-Astronomical-Observatory Multi-object Integral-Field Spectrograph) galaxy survey \citep{croom+12:sami,bryant+15:sami}. However, they also found the trend of higher-mass galaxies having larger triaxialities, in agreement with our results and those of the other studies.

\begin{figure*}
    \centering
    \includegraphics{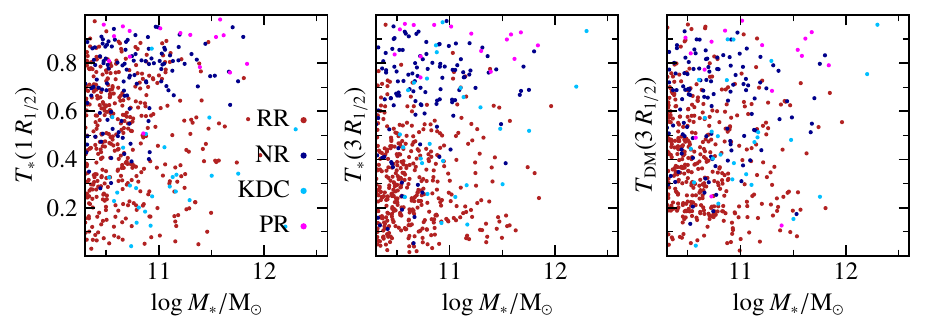}
    \caption{Relation between triaxiality and stellar mass, where the triaxiality is measured for the stellar component at \SI{1}{\Rhalf} and \SI{3}{\Rhalf} (left two panels) and for the DM at \SI{3}{\Rhalf} (right panel), colored by their kinematic class according to the coloring scheme of \citet{schulze+18}. The kinematic classes are regular rotators (RR), non-rotators (NR), kinematically distinct cores (KDC), and prolate rotators (PR).}
    \label{fig:mass_stellar_kinclass}
\end{figure*}

In \cref{fig:mass_stellar_kinclass} we further show where the galaxies of different kinematic classes lie with respect to the relations between the stellar and DM triaxialities and the stellar mass.
Regular rotators are found to have low stellar triaxialities at \SI{3}{\Rhalf} of $T \lesssim 0.5$ (middle panel), while their scatter is much larger for the inner stellar triaxiality and the DM triaxiality.
Non-rotators have relatively large triaxialities for all three considered components, making them overall triaxial to prolate in shape. The scatter in triaxiality is the smallest for $T_*(\SI{1}{\Rhalf})$ (left panel), where more massive non-rotators are always prolate, though this will need to be investigated further to test whether this is an effect of low number statistics.
Both regular rotators and non-rotators have stellar triaxialities consistent with those of the fast and slow rotators presented by \citet{thater+23} for a sample of \atlas{} galaxies, which they obtained through Schwarzschild dynamical modeling. The kinematic classifications they used are the ones from \citet{krajnovic+11:atlas3dII}. Neither the simulations nor the observations indicate a clear relation of the triaxialities with stellar mass.

Kinematically distinct cores mostly have an oblate stellar component in the inner regions (left panel), which corresponds to a disk-like inner shape, consistent with an inner distinctly rotating core. In contrast, the scatter of DM and stellar triaxiality is large at \SI{3}{\Rhalf}, outside of the core region. This large scatter coincides with the overall triaxial shapes that \citet{thater+23} find for the kinematically distinct core galaxies from \atlas{} at one effective radius, also having a relatively large scatter. As the kinematic classification of the \atlas{} galaxies was only based on roughly the inner effective radius, this means that the distinct cores are found well within this radius, whereas the distinct cores classified in the simulation can be closer to \SI{1}{\Rhalf} in size, thus leading to more oblate shapes at \SI{1}{\Rhalf} than the \atlas{} distinct cores at \SI{1}{\Reff}.
Finally, prolate rotators have overall very prolate shapes in stellar and DM triaxialities at both considered radii.

\begin{figure*}
    \centering
    \includegraphics{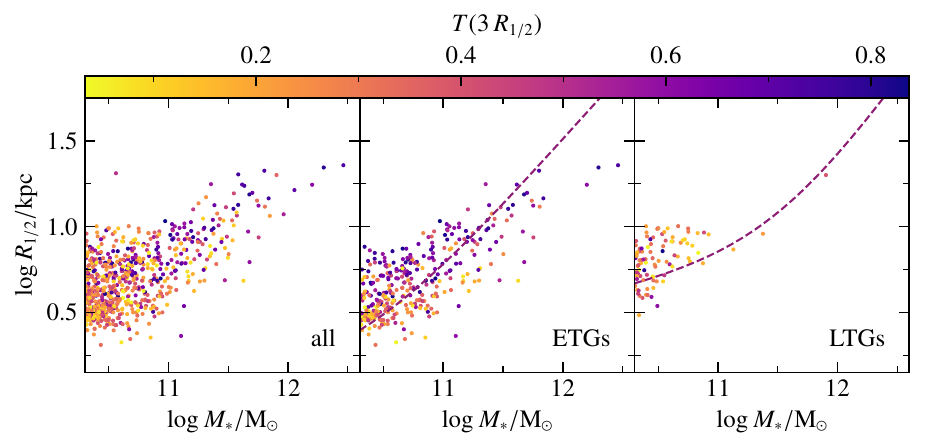}
    \caption{Mass-size relation of the Magneticum galaxy sample for all galaxies, only ETGs ($b \leq \num{-4.73}$), and only LTGs ($b \geq \num{-4.35}$), colored by triaxiality. The dashed lines indicate the two-component fits to the GAMA ETGs and LTGs by \citet{lange+15} in the $r$-band for their morphology cut.}
    \label{fig:mass_size}
\end{figure*}

\begin{figure*}
    \centering
    \includegraphics{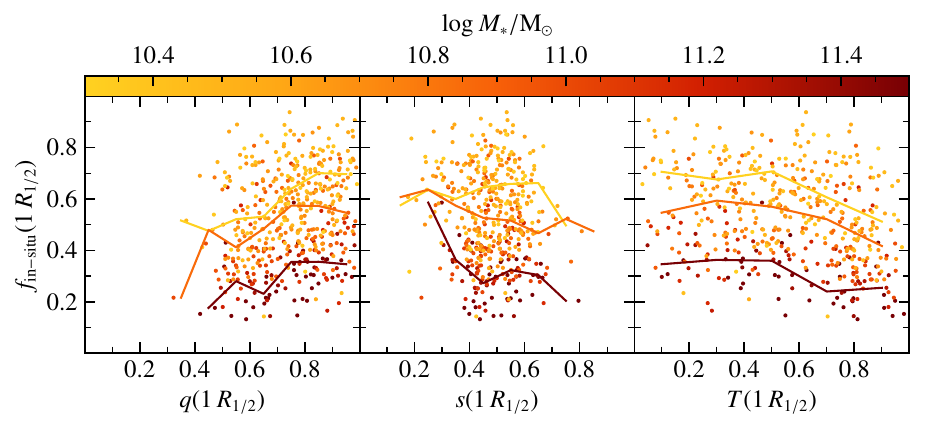}
    \caption{Relation between the in-situ formed stellar fraction within \SI{1}{\Rhalf} and the stellar shape parameters at \SI{1}{\Rhalf}, $q$, $s$, and $T$, colored by the stellar mass. The solid lines indicate the median values in the respective in-situ fraction bins ($\log M_* / \si{\Msun} < \num{10.5}$, $\num{10.5} \leq \log M_* / \si{\Msun} < \num{11.2}$, and $\log M_* / \si{\Msun} \geq \num{11.2}$).}
    \label{fig:insitu_fraction}
\end{figure*}

While the shape is not related to the stellar mass itself, coloring the galaxies by the triaxiality in the mass-size-plane shows that the triaxiality and the relative size of a galaxy compared to other galaxies of the same mass are related (left panel of \cref{fig:mass_size}). Splitting the galaxies in ETGs and LTGs reveals that this trend is valid only for the ETGs (middle panel): ETGs that are large compared to ETGs of the same mass have larger triaxialities, whereas more compact ETGs have small values of $T$. For the intermediate galaxies and LTGs, no such trend can be found (right panel for the LTGs).

Together with the lack of a correlation with the stellar mass of the shape parameters, this shows that the shapes are not related to the mass itself, but to the \emph{mass distribution} of the stellar component. The difference between ETGs and LTGs may be caused by ETGs having more diverse formation histories, especially with regard to the mass ratios and frequencies of merger events, leading to a clearer relation between the size and shape of the galaxies.

The fraction of in-situ formed stars in a galaxy is a crucial indicator of how violent the galaxy's formation history was \citep[e.g.,][]{karademir+19,remus&forbes22}. Assuming that the shape of a galaxy mainly results from its formation history, we expect there may be a correlation between the shape and in-situ fraction. Considering both quantities within \SI{1}{\Rhalf}, there is a clear relation between all three shape parameters and the in-situ fraction at constant stellar mass (\cref{fig:insitu_fraction}): more circular galaxies from a face-on projection (i.e., larger $q$) and more oblate galaxies tend to have a higher in-situ fraction at constant $M_*$. There is also a less tight trend with $s$, where more spherical galaxies tend to have a smaller in-situ fraction. The trend is the strongest for the triaxiality, which is the most difficult of the three shape parameters to observe, while the more easily obtainable axis ratio $s$ shows the weakest correlation, unfortunately.

It has previously been shown that the stellar mass is strongly related to the in-situ fraction (e.g., \citealp{pillepich+14,pillepich+18:tng} for IllustrisTNG; \citealp{remus&forbes22} for Magneticum Box4), where more massive systems have smaller in-situ fractions. This relation can be disentangled even further with the shape parameters. By combining the measured projected shape with an estimate for the inclination of a galaxy and its stellar mass, it is therefore possible to obtain an estimate of the in-situ fraction, a quantity that is currently not possible to directly measure.

\subsection{Dark matter component}
\label{sec:dm_global_properties}

\begin{figure}
    \centering
    \includegraphics{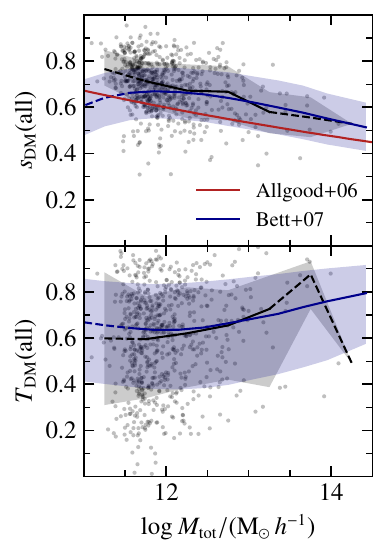}
    \caption{Relation between the total mass and the full DM halo shape parameters, $s_\mathrm{DM}$ and $T_\mathrm{DM}$, for the Magneticum sample used in this work (black), a set of six DM-only simulations \citep[red,][]{allgood+06}, and the Millennium simulation (blue, \citealp{bett+07}). The solid lines indicate the median values in the respective mass bins and the shaded regions the $1\sigma$ ranges (containing \SI{68}{\percent} of the galaxies above and below the median).
    The blue dashed lines mark the halo masses corresponding to less than 300~DM particles in the Millenium simulation. The dashed part of the black lines indicates where there may be effects caused by low number statistics.}
    \label{fig:mass_tot}
\end{figure}

\begin{figure*}
    \centering
    \includegraphics{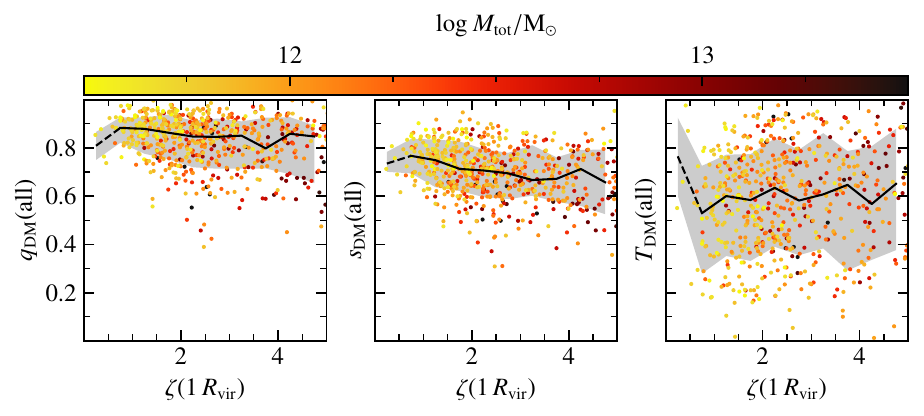}
    \caption{
        Relation between the shape parameters $q$, $s$, and $T$ of the entire DM halos and the gas inflow field asphericity $\zeta$ at a radius of \SI{1}{\Rvir}, where higher $\zeta$ means that the flow field is more aspherical. The individual data points are colored according to the total mass. The black lines indicate the median values in the respective asphericity bins and the shaded regions the $1\sigma$ ranges (containing \SI{68}{\percent} of the galaxies above and below the median). The dashed part of the lines indicates where there may be effects caused by low number statistics.
    }
    \label{fig:dm_asphericity}
\end{figure*}

For the DM component, there have been multiple studies comparing the DM halo shapes with the total halo mass, which overall agree well with each other. An overview of this was presented by \citet{bonamigo+15}, who compared multiple works that studied the halo axis ratio $s$ in DM-only simulations. \Cref{fig:mass_tot} shows that the hydrodynamical Magneticum simulation also features values of $s_\mathrm{DM}$ and $T_\mathrm{DM}$ for the total DM halos that agree with previous findings: In the analyzed total mass range, more massive DM halos are flatter than less massive ones, that is the edge-on axis ratio $s$ decreases with total mass, roughly linearly from $s_\mathrm{DM}(\mathrm{all})=0.75$ at $M_\mathrm{tot} = \SI[parse-numbers=false]{10^{11.5}}{\Msun}$ down to $s_\mathrm{DM}(\mathrm{all})=0.5$ at $M_\mathrm{tot} = \SI{e14}{\Msun}$ (black line). The best agreement is found with the study by \citet{bett+07} for the DM-only Millenium simulation (blue line), who also obtained a similar scatter of values. \Citet{allgood+06} found a similar downward slope for a set of six DM-only simulations, although with slightly flatter halos, which have lower values of $s$ (red line). However, they used a different definition of a halo, determining the shapes at a radius of \SI{0.3}{\Rvir} instead of for all the particles corresponding to the given identified halo. Overall, the DM halos found in Magneticum at the lower mass end ($M_\mathrm{tot} \lesssim \SI{e12}{\Msun}$) are slightly more spherical than those found by DM-only simulations. This is likely a result of the inclusion of baryonic matter, which was found by \citet{chua+19} to lead to rounder DM halos in the Illustris simulation \citep{vogelsberger+14:illustris}. However, in the higher-mass range of groups and clusters, we cannot confirm this trend as it appears that the few more massive systems have shapes more closely resembling those of the DM-only simulations. This would be consistent with the idea that the large-scale structures are still mainly dominated by DM, whereas the baryonic component plays an increasingly strong role on smaller scales.
However, note that at this mass range we are limited by low number statistics as the box volume does not contain many groups and clusters.

The bottom panel of \cref{fig:mass_tot} shows that the triaxiality of the DM halos increases to more prolate shapes for more massive halos, although the scatter is much larger than for $s$ (black line). Both the median relation between total mass and triaxiality and the scatter are again consistent with the findings of \citet{bett+07} for the DM-only Millenium simulation (blue line). The slight deviation of the slope of the relation is negligible compared to the large scatter. This finding at first stands in contrast to the conclusion that \citet{chua+19} drew, which is that including baryonic matter in the simulations leads to more oblate halos. However, as they also found that the systems tend to be more spherical when including baryonic matter, this also means that small fluctuations lead to large changes of the triaxiality, therefore making it more difficult to constrain the triaxialities. The only mass range where we also find slightly more oblate shapes compared to \citet{bett+07} is at $M_\mathrm{tot} \lesssim \SI{e12}{\Msun}$, the same regime as the one where we also found more spherical systems than them in the top panel of the figure. Again, this is consistent with the picture that baryonic matter affects the smaller scales more strongly than the large scales.

In a recent study modeling the shapes of brightest cluster galaxies (BCGs) in observations and comparing their values to Magneticum Box4 (uhr) and TNG100, \citet{de_nicola+22:shapes} found overall similar values of the halo axis ratios for galaxies with $M_\mathrm{tot} > \SI{e13}{\Msun}$, especially for the more massive systems. For the lower-mass systems, they found slightly flatter halos than the simulated galaxies, being even flatter than the more massive halos they modeled. This would contradict the overall trend found from cosmological simulations, though possible causes for this deviation are low number statistics, the uncertainties of the deprojection method applied to the observed BCGs, or differences between the DM and stellar halo shapes at large radii.

While it has been found that there is also a likely trend with the cosmic web environment \citep[e.g.,][]{ganeshaiah_veena+18}, all studies including this one have found that the DM halo axis ratio $s_\mathrm{DM}$ decreases with rising halo mass and the halos become more prolate.
\Citet{bett+07} also attributed the more prolate halo shapes to more massive galaxies accreting matter from large-scale filaments, leading to more elongated shapes.
We tested this by determining the gas inflow field asphericity $\zeta$ at a radius of \SI{1}{\Rvir} (\cref{sec:asphericity}). This parameter quantifies the anisotropy of the gas inflows from the larger scale, indicating how strongly a galaxy is fed by filaments and other anisotropic large-scale structures (for a more detailed analysis of the asphericity, see Kimmig et al.\ in prep.; Seidel et al.\ in prep.). We then compared the DM halo shapes of our simulated galaxy sample to the gas inflow asphericity (\cref{fig:dm_asphericity}). We found that while the face-on axis ratio $q$ and the triaxiality $T$ do not correlate with the asphericity, the halos are flatter (smaller values of $s$) when the gas inflows are more anisotropic. This means that galaxies with stronger filamentary inflows tend to also be less spherical in shape. As the asphericity is also correlated with the total mass, where more massive halos tend to have less isotropic inflows (data point colors), this also explains the overall trend for lower values of $s_\mathrm{DM}$ with higher total mass seen in \cref{fig:mass_tot}.
Thus, we conclude that the halo shapes at large radii are in fact correlated with the filamentarity of the infall.

\section{Shape alignment and dynamics}
\label{sec:alignment}

\begin{figure*}
    \centering
    \includegraphics{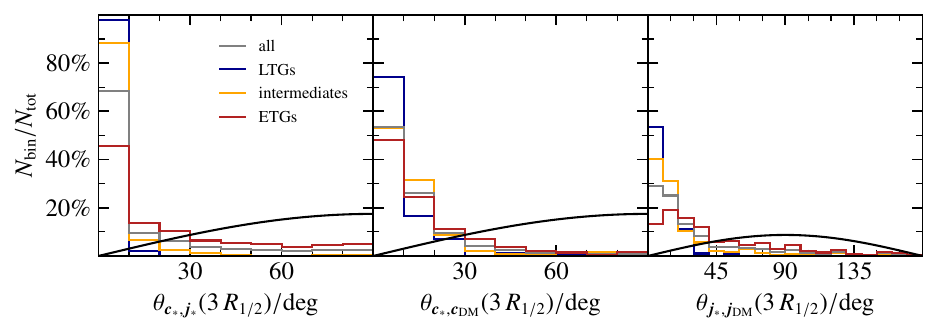}
    \caption{Alignment angles for all galaxies in the sample and for the morphological classes. All histograms are binned with bin widths of \SI{10}{\degree}. The black lines denote the distribution of alignment angles for randomly oriented shapes and angular momenta.
    \emph{Left}: Angle between the stellar minor axis of the shape and the stellar angular momentum within \SI{3}{\Rhalf}.
    \emph{Middle}: Angle between the stellar and DM minor axes of the shapes at \SI{3}{\Rhalf}.
    \emph{Right}: Angle between the stellar and DM angular momenta within \SI{3}{\Rhalf} (unlike the other two angles, this can also reach values up to \SI{180}{\degree}).}
    \label{fig:alignment_histogram}
\end{figure*}

Having addressed the relations between the shape parameters themselves and other galaxy properties, the question remains which properties the \emph{orientation} of the 3D ellipsoids typically corresponds to. A reasonable assumption is that the shapes are aligned with the angular momentum. The left panel of \cref{fig:alignment_histogram} shows the distribution of alignment angles between the minor axis of the stellar shape ellipsoid, $\vec{c}_*$, and the stellar angular momentum, $\vec{j}_*$, within \SI{3}{\Rhalf}. For all morphological classes, a large majority of the galaxies have a strong alignment between the shape and the galaxy's motion, as expected. The alignment is the strongest for LTGs (\SI{97}{\percent} are aligned within \SI{10}{\degree}), closely followed by intermediate galaxies (\SI{88}{\percent}). This is consistent with the idea that these typically flat and oblate galaxies, which are more rotationally dominated (\cref{fig:lambda_r_bvalue}), would rotate within the plane given by their flattened disk-like shape. In contrast, ETGs are only half as likely to feature such an alignment of less than \SI{10}{\degree} (\SI{46}{\percent}) and can even feature prolate rotation with misalignments of close to \SI{90}{\degree}. Around \SI{5}{\percent} of ETGs have misaligned angular momenta by ${\geq}\SI{80}{\degree}$ and \SI{10}{\percent} by ${\geq}\SI{70}{\degree}$. The properties of the shapes of prolate rotators and their alignments will be discussed in more detail in a forthcoming paper (Valenzuela et al.\ in prep.).

Having found a dependency between the stellar and DM shapes at \SI{3}{\Rhalf} (\cref{fig:shapes_stellar_dm}), a second alignment can be expected between the orientations of the stellar and DM shape ellipsoids. The middle panel of \cref{fig:alignment_histogram} shows that indeed most galaxies have aligning shapes, quantified by the angle between the minor axes, $\vec{c}_*$ and $\vec{c}_\mathrm{DM}$. Like for the alignment between the stellar shape ellipsoid and the angular momentum, the alignment between the stellar and DM shapes is the strongest for LTGs, followed by intermediate galaxies and finally by ETGs. For LTGs, near-perfect alignment is less common than for the angular momentum: \SI{77}{\percent} of LTGs have stellar and DM shapes aligned by less than \SI{10}{\degree}. Such an alignment is also less common for intermediate galaxies (\SI{52}{\percent}), but the same for ETGs (\SI{46}{\percent}).

While LTGs and intermediate galaxies have larger average shape alignment angles between the stellar and DM components than the shape alignment angles with the angular momentum, it is the opposite for ETGs: there are fewer ETGs that have large misalignments between the stellar and DM shapes, with \SI{83}{\percent} having $\theta_{\vec{c}_*,\vec{c}_\mathrm{DM}} \leq \SI{30}{\degree}$.
Together with \cref{fig:shapes_stellar_dm}, these findings indicate that the DM follows the stellar component both in shape and orientation.

To test how dynamically aligned the stellar and DM components are, we compared the orientations of the stellar and DM angular momenta within \SI{3}{\Rhalf} with each other (right panel of \cref{fig:alignment_histogram}). An alignment of below \SI{10}{\degree} is less common than the previous two alignment angles we considered: this is the case for \SI{48}{\percent} of the LTGS, \SI{40}{\percent} of the intermediate galaxies, and only \SI{13}{\percent} of the ETGs. Interestingly, the most likely alignment of ETGs is not below \SI{10}{\degree}, but between \SIlist{10;20}{\degree}, although the misalignment can reach large values, especially compared to the other two morphological classes.
These results show that while the DM can adapt its shape to more closely follow the stellar component, its angular momentum cannot be easily redistributed, therefore leading to larger misalignments compared to its shape orientation.
The relation between the absolute stellar and DM angular momenta and their alignments with each other are presented in \cref{app:angular_momenta}.

\begin{figure*}
    \centering
    \includegraphics{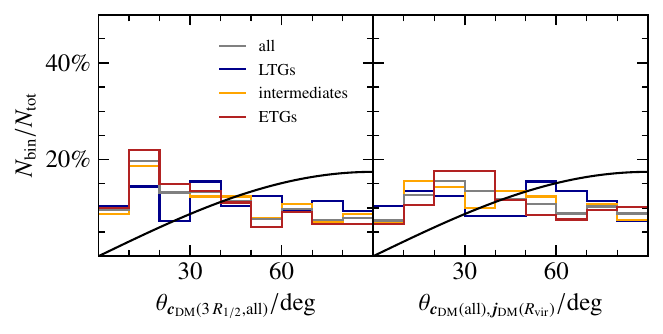}
    \caption{Alignment angles for the galaxies in the sample and their halos, subdivided by the morphological classes. Both histograms are binned with bin widths of \SI{10}{\degree}. The black lines denote the distribution of alignment angles for randomly oriented shapes and angular momenta.
    \emph{Left}: Angle between the minor axes of the DM shapes at \SI{3}{\Rhalf} and of the entire DM halo (all).
    \emph{Right}: Angle between the minor axis of the DM halo and its angular momentum within \SI{1}{\Rvir}.}
    \label{fig:alignment_histogram_dm_all}
\end{figure*}

Finally, we investigated whether the orientation of the inner DM shape at \SI{3}{\Rhalf} aligns with that of the total DM halo. The left panel of \cref{fig:alignment_histogram_dm_all} shows that the alignment angles are very spread out for all galaxies with no significant differences between the morphological types (colored lines). They are still more aligned than randomly oriented shapes (black line), which shows that there is a fine link between the inner and outer regions of galaxies, though for any practical purposes, the distribution of alignment angles does not help constrain these.
This agrees well with our finding that the inner DM shape is largely independent of the halo shape with respect to the axis ratios and triaxiality (\cref{sec:dm_shapes}).
\Citet{xu+23:tng} investigate the galaxy-halo alignment in TNG300 and find median alignment angles between \SI{0.35}{\degree} and~\SI{0.55}{\degree}, with lower-mass galaxies being more aligned with their halos. These values are higher than the ones we find for Magneticum galaxies between the inner DM halo at \SI{3}{\Rhalf} and the overall halo as they compared the DM halo to the stellar orientation at \SI{2}{\Rhalf}, that is for a different particle type and at a smaller radius.

As the shape of the inner stellar component is well aligned with its own angular momentum, we tested whether the DM halo shape is also aligned with the halo angular momentum. The right panel of \cref{fig:alignment_histogram_dm_all} shows that in fact the minor axis of the halo is similarly poorly aligned with the halo angular momentum measured within \SI{1}{\Rvir} as it is with the inner DM shape. Again, the distribution of angles is more aligned than the expected distribution for random orientations, and there is no significant difference found for galaxies of different morphological types.
We propose that the reason for this lack of angular momentum alignment is the fact that the halos are typically more spherical and prolate than a galaxy's inner stellar component, which tends to be flatter and more oblate. It is more likely for these spherical and prolate shapes to have rotations along other directions besides only the minor axis.
To conclude, the halo shapes are much more difficult to constrain and relate to other properties than the inner stellar or DM shapes.

\section{Summary and conclusion}
\label{sec:summary_conclusion}

In this work, we presented an analysis of the 3D shapes of the stellar and DM components of galaxies at $z=0.06$ in the Magneticum Pathfinder \MagneticumBox{4}{uhr} simulation and how they relate to other galaxy properties. The galaxy sample consists of 690 main galaxies with stellar masses of $M_* \geq \SI{2e10}{\Msun}$, DM masses of $M_\mathrm{DM} \geq \SI{e11}{\Msun}$, and stellar half-mass radii of $\Rhalf \geq \SI{2}{\kilo\parsec}$. In this work, all substructures are removed to not influence the properties extracted for the main galaxies.
To determine the shapes and kinematic parameters, we shift the positions and velocities of the particles around the center of particles and velocity.
We then computed the shapes through the iterative unweighted shape determination method, for which all particles within an iteratively deformed ellipsoid of constant volume are considered. We identified this method as the best from several different options found throughout the literature, which we tested for our simulated sample of galaxies, thereby extending the work of \citet{zemp+11}.
The alternative methods have the following caveats: The reduced and non-iterative methods are biased towards spherical shapes; the reduced ellipsoidal methods are only useful for obtaining an average shape of a system (and not a local one as we were interested in for this work); and the iterative methods keeping the major axis length constant makes a consistent comparison of equal-mass galaxies more difficult than the constant volume methods.

The shapes are quantified through the ellipsoidal axis ratios $q = b/a$ and $s = c/a$, and through the triaxiality, which describes how oblate, prolate, or triaxial the shape is.
There is a direct connection between the 3D ellipsoid axis ratios and the projected 2D shape for the face-on and edge-on projections. In those cases, $s$ corresponds to the edge-on axis ratio and $q$ to the face-on axis ratio.

For this study, we computed the shapes of the stellar component at \SIlist{1;3}{\Rhalf} and those of the DM at \SI{3}{\Rhalf} and for the entire DM halo.
We found the following key results:
\begin{itemize}
    \item The stellar shapes are strongly correlated with the morphology approximated by the \bvalue{}, and to the amount of rotational support, quantified through the \lambdaR{}-parameter. ETGs are more prolate than LTGs, and LTGs also become more oblate at \SI{3}{\Rhalf} compared to at \SI{1}{\Rhalf}.
    \item The stellar and DM shapes are strongly correlated at \SI{3}{\Rhalf}: their ellipsoidal axis ratios, triaxialities, and orientations align well with each other. The minor axis of the DM even aligns better with that of the stellar component than with its own angular momentum. As the baryonic matter generally is dominant in the inner regions of galaxies, this indicates that the DM follows the baryonic potential with respect to its shape and orientation.
    \item The stellar shape is not related to the stellar mass, but to the distribution of mass, visible in the mass-size relation. This is especially the case for ETGs, where more extended galaxies have larger triaxialities at constant stellar mass.
    \item Regular rotators tend to be oblate, kinematically distinct cores are oblate in their core regions, non-rotators are triaxial to prolate, and prolate rotators are also prolate in shape.
    \item All three shape parameters further constrain the fraction of in-situ formed stars together with the stellar mass, improving the determination of a quantity that is otherwise very difficult to measure. The relation is the clearest when using the triaxiality of the stellar component.
    \item The outer DM halo shape is detached from the inner behavior. Its shape is much more prolate, consistent with previous literature. Thus, care must be taken when modeling the DM shape for Schwarzschild or Jean's modeling, for example. The DM halo shape is more closely related to the asphericity of the gas inflow field at the virial radius, showing that it is influenced by the large-scale properties.
\end{itemize}

This work has shown that the shapes of galaxies can be used as indicators for a large range of galaxy properties, including ones that are intrinsically difficult to measure such as the in-situ formed fraction of stars. These will help place constraints on derived quantities from observations as modeling techniques for obtaining the intrinsic 3D shapes of galaxies and their halos further improve. The tight relations shown with the kinematics and between the stellar and DM components also lay the foundation for future more focused studies of galaxy shapes, which we will address in a forthcoming paper on the radial shape profiles of the inner regions of galaxies and which will have great relevance for improving triaxial Schwarzschild modeling techniques.
Finally, the correlation of the halo shape to the geometry of the large-scale gas inflow field clearly highlights that measuring the shapes at such large radii through tracer populations like globular clusters and planetary nebulae could be a possible way to infer the large-scale structure around a galaxy. Whether this is possible needs to be addressed in more detail in a future study.

\begin{acknowledgements}
We thank Moritz Fischer for helpful discussions on the manuscript.
We also thank the anonymous referee for useful comments that improved the paper.

LMV acknowledges support by the German Academic Scholarship Foundation (Studienstiftung des deutschen Volkes) and the Marianne-Plehn-Program of the Elite Network of Bavaria.
LMV, KD, and BAS acknowledge support by the COMPLEX project from the European Research Council (ERC) under the European Union’s Horizon 2020 research and innovation program grant agreement ERC-2019-AdG 882679.

The calculations for the hydrodynamical simulations were carried out at the Leibniz Supercomputer Center (LRZ) under the project pr83li (Magneticum).

This research was supported by the Excellence Cluster ORIGINS, funded by the Deutsche Forschungsgemeinschaft under Germany's Excellence Strategy – EXC-2094-390783311.

The following software was used for this work: Julia \citep{bezanson+17:julia}, CSV.jl \citep{quinn+:csv.jl}, DataFrames.jl \citep{kaminski+:dataframes.jl}, GadgetIO.jl \citep{boess&valenzuela:gadgetio.jl}, Matplotlib \citep{hunter07:matplotlib}.
\end{acknowledgements}

\bibliographystyle{style/aa}
\bibliography{bib}

\begin{thebibliography}{147}
\expandafter\ifx\csname natexlab\endcsname\relax\def\natexlab#1{#1}\fi

\bibitem[{Aarseth \& Binney(1978)}]{aarseth&binney78}
Aarseth, S.~J. \& Binney, J. 1978, \mnras, 185, 227

\bibitem[{Allgood {et~al.}(2006)Allgood, Flores, Primack, Kravtsov, Wechsler, Faltenbacher, \& Bullock}]{allgood+06}
Allgood, B., Flores, R.~A., Primack, J.~R., {et~al.} 2006, \mnras, 367, 1781

\bibitem[{Arnaboldi {et~al.}(1996)Arnaboldi, Freeman, Méndez, Capaccioli, Ciardullo, Ford, Gerhard, Hui, Jacoby, Kudritzki, \& Quinn}]{arnaboldi+96}
Arnaboldi, M., Freeman, K.~C., Méndez, R.~H., {et~al.} 1996, \apj, 472, 145

\bibitem[{Arnaboldi {et~al.}(2012)Arnaboldi, Ventimiglia, Iodice, Gerhard, \& Coccato}]{arnaboldi+12}
Arnaboldi, M., Ventimiglia, G., Iodice, E., Gerhard, O., \& Coccato, L. 2012, \aap, 545, A37

\bibitem[{Arnold {et~al.}(2014)Arnold, Romanowsky, Brodie, Forbes, Strader, Spitler, Foster, Blom, Kartha, Pastorello, Pota, Usher, \& Woodley}]{arnold+14}
Arnold, J.~A., Romanowsky, A.~J., Brodie, J.~P., {et~al.} 2014, \apj, 791, 80

\bibitem[{Bacon {et~al.}(2001)Bacon, Copin, Monnet, Miller, {Allington-Smith}, Bureau, Carollo, Davies, Emsellem, Kuntschner, Peletier, Verolme, \& {de Zeeuw}}]{bacon+01:sauronI}
Bacon, R., Copin, Y., Monnet, G., {et~al.} 2001, \mnras, 326, 23

\bibitem[{Bailin \& Steinmetz(2005)}]{bailin&steinmetz05}
Bailin, J. \& Steinmetz, M. 2005, \apj, 627, 647

\bibitem[{Barnes(1992)}]{barnes92}
Barnes, J.~E. 1992, \apj, 393, 484

\bibitem[{Bellstedt {et~al.}(2018)Bellstedt, Forbes, Romanowsky, Remus, Stevens, Brodie, Poci, McDermid, Alabi, Chevalier, Adams, {Ferré-Mateu}, Wasserman, \& Pandya}]{bellstedt+18}
Bellstedt, S., Forbes, D.~A., Romanowsky, A.~J., {et~al.} 2018, \mnras, 476, 4543

\bibitem[{Bender(1988)}]{bender88}
Bender, R. 1988, \aap, 193, L7

\bibitem[{Bett(2012)}]{bett12}
Bett, P. 2012, \mnras, 420, 3303

\bibitem[{Bett {et~al.}(2007)Bett, Eke, Frenk, Jenkins, Helly, \& Navarro}]{bett+07}
Bett, P., Eke, V., Frenk, C.~S., {et~al.} 2007, \mnras, 376, 215

\bibitem[{Bezanson {et~al.}(2017)Bezanson, Edelman, Karpinski, \& Shah}]{bezanson+17:julia}
Bezanson, J., Edelman, A., Karpinski, S., \& Shah, V.~B. 2017, SIAM Review, 59, 65

\bibitem[{Binney(1978)}]{binney78}
Binney, J. 1978, \mnras, 183, 501

\bibitem[{Binney(2005)}]{binney05}
Binney, J. 2005, \mnras, 363, 937

\bibitem[{Bonamigo {et~al.}(2015)Bonamigo, Despali, Limousin, Angulo, Giocoli, \& Soucail}]{bonamigo+15}
Bonamigo, M., Despali, G., Limousin, M., {et~al.} 2015, \mnras, 449, 3171

\bibitem[{Brodie {et~al.}(2014)Brodie, Romanowsky, Strader, Forbes, Foster, Jennings, Pastorello, Pota, Usher, Blom, Kader, Roediger, Spitler, Villaume, Arnold, Kartha, \& Woodley}]{brodie+14:sluggs}
Brodie, J.~P., Romanowsky, A.~J., Strader, J., {et~al.} 2014, \apj, 796, 52

\bibitem[{Bryan {et~al.}(2013)Bryan, Kay, Duffy, Schaye, Dalla~Vecchia, \& Booth}]{bryan+13}
Bryan, S.~E., Kay, S.~T., Duffy, A.~R., {et~al.} 2013, \mnras, 429, 3316

\bibitem[{Bryant {et~al.}(2015)Bryant, Owers, Robotham, Croom, Driver, Drinkwater, Lorente, Cortese, Scott, Colless, Schaefer, Taylor, Konstantopoulos, Allen, Baldry, Barnes, Bauer, {Bland-Hawthorn}, Bloom, Brooks, Brough, Cecil, Couch, Croton, Davies, Ellis, Fogarty, Foster, Glazebrook, Goodwin, Green, Gunawardhana, Hampton, Ho, Hopkins, Kewley, Lawrence, {Leon-Saval}, Leslie, McElroy, Lewis, Liske, {López-Sánchez}, Mahajan, Medling, Metcalfe, Meyer, Mould, Obreschkow, O'Toole, Pracy, Richards, Shanks, Sharp, Sweet, Thomas, Tonini, \& Walcher}]{bryant+15:sami}
Bryant, J.~J., Owers, M.~S., Robotham, A. S.~G., {et~al.} 2015, \mnras, 447, 2857

\bibitem[{Bundy {et~al.}(2015)Bundy, Bershady, Law, Yan, Drory, MacDonald, Wake, Cherinka, {Sánchez-Gallego}, Weijmans, Thomas, Tremonti, Masters, Coccato, {Diamond-Stanic}, {Aragón-Salamanca}, {Avila-Reese}, Badenes, {Falcón-Barroso}, Belfiore, Bizyaev, Blanc, {Bland-Hawthorn}, Blanton, Brownstein, Byler, Cappellari, Conroy, Dutton, Emsellem, Etherington, Frinchaboy, Fu, Gunn, Harding, Johnston, Kauffmann, Kinemuchi, Klaene, Knapen, Leauthaud, Li, Lin, Maiolino, Malanushenko, Malanushenko, Mao, Maraston, McDermid, Merrifield, Nichol, Oravetz, Pan, Parejko, Sanchez, Schlegel, Simmons, Steele, Steinmetz, Thanjavur, Thompson, Tinker, {van den Bosch}, Westfall, Wilkinson, Wright, Xiao, \& Zhang}]{bundy+15:manga}
Bundy, K., Bershady, M.~A., Law, D.~R., {et~al.} 2015, \apj, 798, 7

\bibitem[{Butsky {et~al.}(2016)Butsky, Macciò, Dutton, Wang, Obreja, Stinson, Penzo, Kang, Keller, \& Wadsley}]{butsky+16}
Butsky, I., Macciò, A.~V., Dutton, A.~A., {et~al.} 2016, \mnras, 462, 663

\bibitem[{Bílek {et~al.}(2020)Bílek, Duc, Cuillandre, Gwyn, Cappellari, Bekaert, Bonfini, Bitsakis, Paudel, Krajnović, Durrell, \& Marleau}]{bilek+20}
Bílek, M., Duc, P.-A., Cuillandre, J.-C., {et~al.} 2020, \mnras, 498, 2138

\bibitem[{Böss \& Valenzuela(2023)}]{boess&valenzuela:gadgetio.jl}
Böss, L.~M. \& Valenzuela, L.~M. 2023, LudwigBoess/GadgetIO.Jl

\bibitem[{Cappellari(2008)}]{cappellari08:jam}
Cappellari, M. 2008, \mnras, 390, 71

\bibitem[{Cappellari \& Copin(2003)}]{cappellari&copin03:cvt}
Cappellari, M. \& Copin, Y. 2003, \mnras, 342, 345

\bibitem[{Cappellari {et~al.}(2007)Cappellari, Emsellem, Bacon, Bureau, Davies, {de Zeeuw}, {Falcón-Barroso}, Krajnović, Kuntschner, McDermid, Peletier, Sarzi, {van den Bosch}, \& {van de Ven}}]{cappellari+07:sauronX}
Cappellari, M., Emsellem, E., Bacon, R., {et~al.} 2007, \mnras, 379, 418

\bibitem[{Cappellari {et~al.}(2011)Cappellari, Emsellem, Krajnović, McDermid, Scott, Verdoes~Kleijn, Young, Alatalo, Bacon, Blitz, Bois, Bournaud, Bureau, Davies, Davis, {de Zeeuw}, Duc, Khochfar, Kuntschner, Lablanche, Morganti, Naab, Oosterloo, Sarzi, Serra, \& Weijmans}]{cappellari+11:atlas3dI}
Cappellari, M., Emsellem, E., Krajnović, D., {et~al.} 2011, \mnras, 413, 813

\bibitem[{Cataldi {et~al.}(2021)Cataldi, Pedrosa, Tissera, \& Artale}]{cataldi+21}
Cataldi, P., Pedrosa, S.~E., Tissera, P.~B., \& Artale, M.~C. 2021, \mnras, 501, 5679

\bibitem[{Chua {et~al.}(2019)Chua, Pillepich, Vogelsberger, \& Hernquist}]{chua+19}
Chua, K. T.~E., Pillepich, A., Vogelsberger, M., \& Hernquist, L. 2019, \mnras, 484, 476

\bibitem[{Croom {et~al.}(2012)Croom, Lawrence, {Bland-Hawthorn}, Bryant, Fogarty, Richards, Goodwin, Farrell, Miziarski, Heald, Jones, Lee, Colless, Brough, Hopkins, Bauer, Birchall, Ellis, Horton, {Leon-Saval}, Lewis, {López-Sánchez}, Min, Trinh, \& Trowland}]{croom+12:sami}
Croom, S.~M., Lawrence, J.~S., {Bland-Hawthorn}, J., {et~al.} 2012, \mnras, 421, 872

\bibitem[{{de Nicola} {et~al.}(2022{\natexlab{a}}){de Nicola}, Neureiter, Thomas, Saglia, \& Bender}]{de_nicola+22}
{de Nicola}, S., Neureiter, B., Thomas, J., Saglia, R.~P., \& Bender, R. 2022{\natexlab{a}}, \mnras, 517, 3445

\bibitem[{{de Nicola} {et~al.}(2022{\natexlab{b}}){de Nicola}, Saglia, Thomas, Pulsoni, Kluge, Bender, Valenzuela, \& Remus}]{de_nicola+22:shapes}
{de Nicola}, S., Saglia, R.~P., Thomas, J., {et~al.} 2022{\natexlab{b}}, \apj, 933, 215

\bibitem[{{de Vaucouleurs}(1959)}]{de_vaucouleurs59}
{de Vaucouleurs}, G. 1959, Handbuch der Physik, 53, 275

\bibitem[{Derkenne {et~al.}(2021)Derkenne, McDermid, Poci, Remus, Jørgensen, \& Emsellem}]{derkenne+21}
Derkenne, C., McDermid, R.~M., Poci, A., {et~al.} 2021, \mnras, 506, 3691

\bibitem[{Dolag {et~al.}(2009)Dolag, Borgani, Murante, \& Springel}]{dolag+09:subfind}
Dolag, K., Borgani, S., Murante, G., \& Springel, V. 2009, \mnras, 399, 497

\bibitem[{Dolag {et~al.}(2005{\natexlab{a}})Dolag, Hansen, Roncarelli, \& Moscardini}]{dolag+05:smac}
Dolag, K., Hansen, F.~K., Roncarelli, M., \& Moscardini, L. 2005{\natexlab{a}}, \mnras, 363, 29

\bibitem[{Dolag {et~al.}(2004)Dolag, Jubelgas, Springel, Borgani, \& Rasia}]{dolag+04}
Dolag, K., Jubelgas, M., Springel, V., Borgani, S., \& Rasia, E. 2004, \apjl, 606, L97

\bibitem[{Dolag {et~al.}(2005{\natexlab{b}})Dolag, Vazza, Brunetti, \& Tormen}]{dolag+05}
Dolag, K., Vazza, F., Brunetti, G., \& Tormen, G. 2005{\natexlab{b}}, \mnras, 364, 753

\bibitem[{Dolfi {et~al.}(2021)Dolfi, Forbes, Couch, Bekki, {Ferré-Mateu}, Romanowsky, \& Brodie}]{dolfi+21}
Dolfi, A., Forbes, D.~A., Couch, W.~J., {et~al.} 2021, \mnras, 504, 4923

\bibitem[{Donnert {et~al.}(2013)Donnert, Dolag, Brunetti, \& Cassano}]{donnert+13}
Donnert, J., Dolag, K., Brunetti, G., \& Cassano, R. 2013, \mnras, 429, 3564

\bibitem[{Dubinski \& Carlberg(1991)}]{dubinski&carlberg91}
Dubinski, J. \& Carlberg, R.~G. 1991, \apj, 378, 496

\bibitem[{Duc {et~al.}(2015)Duc, Cuillandre, Karabal, Cappellari, Alatalo, Blitz, Bournaud, Bureau, Crocker, Davies, Davis, {de Zeeuw}, Emsellem, Khochfar, Krajnović, Kuntschner, McDermid, {Michel-Dansac}, Morganti, Naab, Oosterloo, Paudel, Sarzi, Scott, Serra, Weijmans, \& Young}]{duc+15:atlas3dXXIX}
Duc, P.-A., Cuillandre, J.-C., Karabal, E., {et~al.} 2015, \mnras, 446, 120

\bibitem[{Emami {et~al.}(2021)Emami, Hernquist, Alcock, Genel, Bose, Weinberger, Vogelsberger, Shen, Speagle, Marinacci, Forbes, \& Torrey}]{emami+21}
Emami, R., Hernquist, L., Alcock, C., {et~al.} 2021, \apj, 918, 7

\bibitem[{Emsellem {et~al.}(2011)Emsellem, Cappellari, Krajnović, Alatalo, Blitz, Bois, Bournaud, Bureau, Davies, Davis, {de Zeeuw}, Khochfar, Kuntschner, Lablanche, McDermid, Morganti, Naab, Oosterloo, Sarzi, Scott, Serra, {van de Ven}, Weijmans, \& Young}]{emsellem+11:atlas3dIII}
Emsellem, E., Cappellari, M., Krajnović, D., {et~al.} 2011, \mnras, 414, 888

\bibitem[{Emsellem {et~al.}(2007)Emsellem, Cappellari, Krajnović, {van de Ven}, Bacon, Bureau, Davies, {de Zeeuw}, {Falcón-Barroso}, Kuntschner, McDermid, Peletier, \& Sarzi}]{emsellem+07:sauronIX}
Emsellem, E., Cappellari, M., Krajnović, D., {et~al.} 2007, \mnras, 379, 401

\bibitem[{Fabjan {et~al.}(2010)Fabjan, Borgani, Tornatore, Saro, Murante, \& Dolag}]{fabjan+10}
Fabjan, D., Borgani, S., Tornatore, L., {et~al.} 2010, \mnras, 401, 1670

\bibitem[{Fall(1983)}]{fall83}
Fall, S.~M. 1983, in Internal Kinematics and Dynamics of Galaxies, Vol. 100, 391--398

\bibitem[{Fall \& Romanowsky(2013)}]{fall&romanowsky13}
Fall, S.~M. \& Romanowsky, A.~J. 2013, \apjl, 769, L26

\bibitem[{Fischer \& Valenzuela(2023)}]{fischer&valenzuela23}
Fischer, M.~S. \& Valenzuela, L.~M. 2023, \aap, 670, A120

\bibitem[{{Forero-Romero} {et~al.}(2014){Forero-Romero}, Contreras, \& Padilla}]{forero_romero+14}
{Forero-Romero}, J.~E., Contreras, S., \& Padilla, N. 2014, \mnras, 443, 1090

\bibitem[{Foster {et~al.}(2016)Foster, Pastorello, Roediger, Brodie, Forbes, Kartha, Pota, Romanowsky, Spitler, Strader, Usher, \& Arnold}]{foster+16}
Foster, C., Pastorello, N., Roediger, J., {et~al.} 2016, \mnras, 457, 147

\bibitem[{Franx {et~al.}(1991)Franx, Illingworth, \& {de Zeeuw}}]{franx+91}
Franx, M., Illingworth, G., \& {de Zeeuw}, T. 1991, \apj, 383, 112

\bibitem[{Förster {et~al.}(2022)Förster, Remus, Dolag, Kimmig, Teklu, \& Valenzuela}]{foerster+22}
Förster, P.~U., Remus, R.-S., Dolag, K., {et~al.} 2022, arXiv e-prints, arXiv2208.05496

\bibitem[{Ganeshaiah~Veena {et~al.}(2018)Ganeshaiah~Veena, Cautun, {van de Weygaert}, Tempel, Jones, Rieder, \& Frenk}]{ganeshaiah_veena+18}
Ganeshaiah~Veena, P., Cautun, M., {van de Weygaert}, R., {et~al.} 2018, \mnras, 481, 414

\bibitem[{Gebhardt {et~al.}(2003)Gebhardt, Richstone, Tremaine, Lauer, Bender, Bower, Dressler, Faber, Filippenko, Green, Grillmair, Ho, Kormendy, Magorrian, \& Pinkney}]{gebhardt+03}
Gebhardt, K., Richstone, D., Tremaine, S., {et~al.} 2003, \apj, 583, 92

\bibitem[{Gouin {et~al.}(2022)Gouin, Gallo, \& Aghanim}]{gouin+22}
Gouin, C., Gallo, S., \& Aghanim, N. 2022, \aap, 664, A198

\bibitem[{Górski {et~al.}(2005)Górski, Hivon, Banday, Wandelt, Hansen, Reinecke, \& Bartelmann}]{gorski+05:healpix}
Górski, K.~M., Hivon, E., Banday, A.~J., {et~al.} 2005, \apj, 622, 759

\bibitem[{Harris {et~al.}(2020)Harris, Remus, Harris, \& Babyk}]{harris+20}
Harris, W.~E., Remus, R.-S., Harris, G. L.~H., \& Babyk, {\relax Iu}.~V. 2020, \apj, 905, 28

\bibitem[{Hellwing {et~al.}(2021)Hellwing, Cautun, {van de Weygaert}, \& Jones}]{hellwing+21}
Hellwing, W.~A., Cautun, M., {van de Weygaert}, R., \& Jones, B.~T. 2021, \prd, 103, 063517

\bibitem[{Hirschmann {et~al.}(2014)Hirschmann, Dolag, Saro, Bachmann, Borgani, \& Burkert}]{hirschmann+14}
Hirschmann, M., Dolag, K., Saro, A., {et~al.} 2014, \mnras, 442, 2304

\bibitem[{Hohl \& Zang(1979)}]{hohl&zang79}
Hohl, F. \& Zang, T.~A. 1979, \aj, 84, 585

\bibitem[{Hubble(1926)}]{hubble26}
Hubble, E.~P. 1926, \apj, 64, 321

\bibitem[{Hubble(1936)}]{hubble36}
Hubble, E.~P. 1936, Realm of the Nebulae (Yale University Press)

\bibitem[{Hui {et~al.}(1995)Hui, Ford, Freeman, \& Dopita}]{hui+95}
Hui, X., Ford, H.~C., Freeman, K.~C., \& Dopita, M.~A. 1995, \apj, 449, 592

\bibitem[{Hunter(2007)}]{hunter07:matplotlib}
Hunter, J.~D. 2007, Computing in Science and Engineering, 9, 90

\bibitem[{Illingworth(1977)}]{illingworth77}
Illingworth, G. 1977, \apjl, 218, L43

\bibitem[{Jesseit {et~al.}(2009)Jesseit, Cappellari, Naab, Emsellem, \& Burkert}]{jesseit+09}
Jesseit, R., Cappellari, M., Naab, T., Emsellem, E., \& Burkert, A. 2009, \mnras, 397, 1202

\bibitem[{Jethwa {et~al.}(2020)Jethwa, Thater, Maindl, \& {Van de Ven}}]{jethwa+20:dynamite}
Jethwa, P., Thater, S., Maindl, T., \& {Van de Ven}, G. 2020, Astrophysics Source Code Library, ascl:2011.007

\bibitem[{Jin {et~al.}(2020)Jin, Zhu, Long, Mao, Wang, \& {van de Ven}}]{jin+20}
Jin, Y., Zhu, L., Long, R.~J., {et~al.} 2020, \mnras, 491, 1690

\bibitem[{Jing \& Suto(2002)}]{jing&suto02}
Jing, Y.~P. \& Suto, Y. 2002, \apj, 574, 538

\bibitem[{Kamiński {et~al.}(2023)Kamiński, Myles~White, {Powerdistribution}, {Bouchet-Valat}, Garborg, Quinn, Kornblith, {Cjprybol}, Stukalov, Bates, Short, DuBois, Harris, Squire, {Pdeffebach}, Arslan, Anthoff, Kleinschmidt, Noack, Shah, Mellnik, Arakaki, Mohapatra, {Peter}, Karpinski, Lin, {Timema}, {ExpandingMan}, Oswald, \& Arraes Jardim~Chagas}]{kaminski+:dataframes.jl}
Kamiński, B., Myles~White, J., {Powerdistribution}, {et~al.} 2023, JuliaData/DataFrames.Jl

\bibitem[{Karademir {et~al.}(2019)Karademir, Remus, Burkert, Dolag, Hoffmann, Moster, Steinwandel, \& Zhang}]{karademir+19}
Karademir, G.~S., Remus, R.-S., Burkert, A., {et~al.} 2019, \mnras, 487, 318

\bibitem[{Knebe {et~al.}(2010)Knebe, Libeskind, Knollmann, Yepes, Gottlöber, \& Hoffman}]{knebe+10}
Knebe, A., Libeskind, N.~I., Knollmann, S.~R., {et~al.} 2010, \mnras, 405, 1119

\bibitem[{Komatsu {et~al.}(2011)Komatsu, Smith, Dunkley, Bennett, Gold, Hinshaw, Jarosik, Larson, Nolta, Page, Spergel, Halpern, Hill, Kogut, Limon, Meyer, Odegard, Tucker, Weiland, Wollack, \& Wright}]{komatsu+11:wmap7}
Komatsu, E., Smith, K.~M., Dunkley, J., {et~al.} 2011, \apjs, 192, 18

\bibitem[{Kormendy \& Bender(1996)}]{kormendy&bender96}
Kormendy, J. \& Bender, R. 1996, \apjl, 464, L119

\bibitem[{Krajnović {et~al.}(2005)Krajnović, Cappellari, Emsellem, McDermid, \& {de Zeeuw}}]{krajnovic+05}
Krajnović, D., Cappellari, M., Emsellem, E., McDermid, R.~M., \& {de Zeeuw}, P.~T. 2005, \mnras, 357, 1113

\bibitem[{Krajnović {et~al.}(2011)Krajnović, Emsellem, Cappellari, Alatalo, Blitz, Bois, Bournaud, Bureau, Davies, Davis, {de Zeeuw}, Khochfar, Kuntschner, Lablanche, McDermid, Morganti, Naab, Oosterloo, Sarzi, Scott, Serra, Weijmans, \& Young}]{krajnovic+11:atlas3dII}
Krajnović, D., Emsellem, E., Cappellari, M., {et~al.} 2011, \mnras, 414, 2923

\bibitem[{Lange {et~al.}(2015)Lange, Driver, Robotham, Kelvin, Graham, Alpaslan, Andrews, Baldry, Bamford, {Bland-Hawthorn}, Brough, Cluver, Conselice, Davies, Haeussler, Konstantopoulos, Loveday, Moffett, Norberg, Phillipps, Taylor, {López-Sánchez}, \& Wilkins}]{lange+15}
Lange, R., Driver, S.~P., Robotham, A. S.~G., {et~al.} 2015, \mnras, 447, 2603

\bibitem[{Li {et~al.}(2016)Li, Li, Mao, Xu, Long, \& Emsellem}]{li+16}
Li, H., Li, R., Mao, S., {et~al.} 2016, \mnras, 455, 3680

\bibitem[{Li {et~al.}(2018)Li, Mao, Emsellem, Xu, Springel, \& Krajnović}]{li+18:prolate}
Li, H., Mao, S., Emsellem, E., {et~al.} 2018, \mnras, 473, 1489

\bibitem[{Libeskind {et~al.}(2011)Libeskind, Knebe, Hoffman, Gottlöber, Yepes, \& Steinmetz}]{libeskind+11}
Libeskind, N.~I., Knebe, A., Hoffman, Y., {et~al.} 2011, \mnras, 411, 1525

\bibitem[{Liller(1966)}]{liller66}
Liller, M.~H. 1966, \apj, 146, 28

\bibitem[{Ludlow {et~al.}(2021)Ludlow, Fall, Schaye, \& Obreschkow}]{ludlow+21}
Ludlow, A.~D., Fall, S.~M., Schaye, J., \& Obreschkow, D. 2021, \mnras, 508, 5114

\bibitem[{Ludlow {et~al.}(2023)Ludlow, Fall, Wilkinson, Schaye, \& Obreschkow}]{ludlow+23}
Ludlow, A.~D., Fall, S.~M., Wilkinson, M.~J., Schaye, J., \& Obreschkow, D. 2023, \mnras, 525, 5614

\bibitem[{{Martínez-Delgado} {et~al.}(2023){Martínez-Delgado}, Cooper, Román, Pillepich, Erkal, Pearson, Moustakas, Laporte, Laine, Akhlaghi, Lang, Makarov, Borlaff, Donatiello, Pearson, {Miró-Carretero}, Cuillandre, Domínguez, {Roca-Fàbrega}, Frenk, Schmidt, {Gómez-Flechoso}, Guzman, Libeskind, Dey, Weaver, Schlegel, Myers, \& Valdes}]{martinez_delgado+23}
{Martínez-Delgado}, D., Cooper, A.~P., Román, J., {et~al.} 2023, \aap, 671, A141

\bibitem[{Naab {et~al.}(2014)Naab, Oser, Emsellem, Cappellari, Krajnović, McDermid, Alatalo, Bayet, Blitz, Bois, Bournaud, Bureau, Crocker, Davies, Davis, {de Zeeuw}, Duc, Hirschmann, Johansson, Khochfar, Kuntschner, Morganti, Oosterloo, Sarzi, Scott, Serra, {van de Ven}, Weijmans, \& Young}]{naab+14:atlas3dXXV}
Naab, T., Oser, L., Emsellem, E., {et~al.} 2014, \mnras, 444, 3357

\bibitem[{Navarro {et~al.}(1996)Navarro, Frenk, \& White}]{navarro+96:nfw}
Navarro, J.~F., Frenk, C.~S., \& White, S. D.~M. 1996, \apj, 462, 563

\bibitem[{Neureiter {et~al.}(2021)Neureiter, Thomas, Saglia, Bender, Finozzi, Krukau, Naab, Rantala, \& Frigo}]{neureiter+21:smart}
Neureiter, B., Thomas, J., Saglia, R., {et~al.} 2021, \mnras, 500, 1437

\bibitem[{{Nuñez-Castiñeyra} {et~al.}(2023){Nuñez-Castiñeyra}, Nezri, Mollitor, Devriendt, \& Teyssier}]{nunez_castineyra+23}
{Nuñez-Castiñeyra}, A., Nezri, E., Mollitor, P., Devriendt, J., \& Teyssier, R. 2023, \jcap, 2023, 012

\bibitem[{Peng {et~al.}(2004{\natexlab{a}})Peng, Ford, \& Freeman}]{peng+04:GCsII}
Peng, E.~W., Ford, H.~C., \& Freeman, K.~C. 2004{\natexlab{a}}, \apj, 602, 705

\bibitem[{Peng {et~al.}(2004{\natexlab{b}})Peng, Ford, \& Freeman}]{peng+04:PNe}
Peng, E.~W., Ford, H.~C., \& Freeman, K.~C. 2004{\natexlab{b}}, \apj, 602, 685

\bibitem[{Petit {et~al.}(2023)Petit, Ducourant, Slezak, Sluse, \& Delchambre}]{petit+23}
Petit, Q., Ducourant, C., Slezak, E., Sluse, D., \& Delchambre, L. 2023, \aap, 669, A132

\bibitem[{Pilawa {et~al.}(2022)Pilawa, Liepold, Delgado~Andrade, Walsh, Ma, Quenneville, Greene, \& Blakeslee}]{pilawa+22:massiveXVII}
Pilawa, J.~D., Liepold, C.~M., Delgado~Andrade, S.~C., {et~al.} 2022, \apj, 928, 178

\bibitem[{Pillepich {et~al.}(2018)Pillepich, Nelson, Hernquist, Springel, Pakmor, Torrey, Weinberger, Genel, Naiman, Marinacci, \& Vogelsberger}]{pillepich+18:tng}
Pillepich, A., Nelson, D., Hernquist, L., {et~al.} 2018, \mnras, 475, 648

\bibitem[{Pillepich {et~al.}(2014)Pillepich, Vogelsberger, Deason, {Rodriguez-Gomez}, Genel, Nelson, Torrey, Sales, Marinacci, Springel, Sijacki, \& Hernquist}]{pillepich+14}
Pillepich, A., Vogelsberger, M., Deason, A., {et~al.} 2014, \mnras, 444, 237

\bibitem[{Poci {et~al.}(2021)Poci, McDermid, Lyubenova, Zhu, {van de Ven}, Iodice, Coccato, Pinna, Corsini, {Falcón-Barroso}, Gadotti, Grand, Fahrion, {Martín-Navarro}, Sarzi, Viaene, \& {de Zeeuw}}]{poci+21}
Poci, A., McDermid, R.~M., Lyubenova, M., {et~al.} 2021, \aap, 647, A145

\bibitem[{Poci \& Smith(2022)}]{poci&smith22}
Poci, A. \& Smith, R.~J. 2022, \mnras, 512, 5298

\bibitem[{Pota {et~al.}(2013)Pota, Forbes, Romanowsky, Brodie, Spitler, Strader, Foster, Arnold, Benson, Blom, Hargis, Rhode, \& Usher}]{pota+13}
Pota, V., Forbes, D.~A., Romanowsky, A.~J., {et~al.} 2013, \mnras, 428, 389

\bibitem[{Power {et~al.}(2003)Power, Navarro, Jenkins, Frenk, White, Springel, Stadel, \& Quinn}]{power+03:shrinking_sphere}
Power, C., Navarro, J.~F., Jenkins, A., {et~al.} 2003, \mnras, 338, 14

\bibitem[{Prada {et~al.}(2019)Prada, {Forero-Romero}, Grand, Pakmor, \& Springel}]{prada+19}
Prada, J., {Forero-Romero}, J.~E., Grand, R. J.~J., Pakmor, R., \& Springel, V. 2019, \mnras, 490, 4877

\bibitem[{Pulsoni {et~al.}(2020)Pulsoni, Gerhard, Arnaboldi, Pillepich, Nelson, Hernquist, \& Springel}]{pulsoni+20}
Pulsoni, C., Gerhard, O., Arnaboldi, M., {et~al.} 2020, \aap, 641, A60

\bibitem[{Pulsoni {et~al.}(2021)Pulsoni, Gerhard, Arnaboldi, Pillepich, {Rodriguez-Gomez}, Nelson, Hernquist, \& Springel}]{pulsoni+21}
Pulsoni, C., Gerhard, O., Arnaboldi, M., {et~al.} 2021, \aap, 647, A95

\bibitem[{Quenneville {et~al.}(2022)Quenneville, Liepold, \& Ma}]{quenneville+22}
Quenneville, M.~E., Liepold, C.~M., \& Ma, C.-P. 2022, \apj, 926, 30

\bibitem[{Quinn {et~al.}(2023)Quinn, {Bouchet-Valat}, Kamiński, Newman, Stukalov, Vogt, {cjprybol}, Robinson, Noack, Kelman, Davies, {ExpandingMan}, {Ian}, Piibeleht, Finnegan, {evalparse}, Silberstein, Anthony~Blaom, Lungwitz, König, {de Graaf}, Woodfield, Barton, Aluthge, Saba, Noronha, {kragol}, De~Rosario, Ranocha, \& Butterworth}]{quinn+:csv.jl}
Quinn, J., {Bouchet-Valat}, M., Kamiński, B., {et~al.} 2023, JuliaData/CSV.Jl

\bibitem[{Remus {et~al.}(2017)Remus, Dolag, Naab, Burkert, Hirschmann, Hoffmann, \& Johansson}]{remus+17}
Remus, R.-S., Dolag, K., Naab, T., {et~al.} 2017, \mnras, 464, 3742

\bibitem[{Remus \& Forbes(2022)}]{remus&forbes22}
Remus, R.-S. \& Forbes, D.~A. 2022, \apj, 935, 37

\bibitem[{Richstone \& Tremaine(1985)}]{richstone&tremaine85}
Richstone, D.~O. \& Tremaine, S. 1985, \apj, 296, 370

\bibitem[{Rix {et~al.}(1997)Rix, {de Zeeuw}, Cretton, {van der Marel}, \& Carollo}]{rix+97}
Rix, H.-W., {de Zeeuw}, P.~T., Cretton, N., {van der Marel}, R.~P., \& Carollo, C.~M. 1997, \apj, 488, 702

\bibitem[{Rutherford {et~al.}(2024)Rutherford, {van de Sande}, Croom, Valenzuela, Remus, D'Eugenio, Vaughan, Zovaro, Casura, Barsanti, {Bland-Hawthorn}, Brough, Bryant, Goodwin, Lorente, Oh, \& Ristea}]{rutherford+24}
Rutherford, T.~H., {van de Sande}, J., Croom, S.~M., {et~al.} 2024, \mnras, 529, 810

\bibitem[{Santucci {et~al.}(2022)Santucci, Brough, {van de Sande}, McDermid, {van de Ven}, Zhu, D'Eugenio, {Bland-Hawthorn}, Barsanti, Bryant, Croom, Davies, Green, Lawrence, Lorente, Owers, Poci, Richards, Thater, \& Yi}]{santucci+22}
Santucci, G., Brough, S., {van de Sande}, J., {et~al.} 2022, \apj, 930, 153

\bibitem[{Schneider {et~al.}(2012)Schneider, Frenk, \& Cole}]{schneider+12}
Schneider, M.~D., Frenk, C.~S., \& Cole, S. 2012, \jcap, 2012, 030

\bibitem[{Schulze {et~al.}(2020)Schulze, Remus, Dolag, Bellstedt, Burkert, \& Forbes}]{schulze+20}
Schulze, F., Remus, R.-S., Dolag, K., {et~al.} 2020, \mnras, 493, 3778

\bibitem[{Schulze {et~al.}(2018)Schulze, Remus, Dolag, Burkert, Emsellem, \& {van de Ven}}]{schulze+18}
Schulze, F., Remus, R.-S., Dolag, K., {et~al.} 2018, \mnras, 480, 4636

\bibitem[{Schwarzschild(1979)}]{schwarzschild79:schwarzschild_modeling}
Schwarzschild, M. 1979, \apj, 232, 236

\bibitem[{Schwarzschild(1982)}]{schwarzschild82:schwarzschild_modeling}
Schwarzschild, M. 1982, \apj, 263, 599

\bibitem[{Sola {et~al.}(2022)Sola, Duc, Richards, Paiement, Urbano, Klehammer, Bílek, Cuillandre, Gwyn, \& McConnachie}]{sola+22}
Sola, E., Duc, P.-A., Richards, F., {et~al.} 2022, \aap, 662, A124

\bibitem[{Springel(2005)}]{springel05:gadget2}
Springel, V. 2005, \mnras, 364, 1105

\bibitem[{Springel {et~al.}(2004)Springel, White, \& Hernquist}]{springel+04}
Springel, V., White, S. D.~M., \& Hernquist, L. 2004, in Dark Matter in Galaxies, Vol. 220, 421

\bibitem[{Springel {et~al.}(2001)Springel, White, Tormen, \& Kauffmann}]{springel+01:subfind}
Springel, V., White, S. D.~M., Tormen, G., \& Kauffmann, G. 2001, \mnras, 328, 726

\bibitem[{Teklu {et~al.}(2015)Teklu, Remus, Dolag, Beck, Burkert, Schmidt, Schulze, \& Steinborn}]{teklu+15}
Teklu, A.~F., Remus, R.-S., Dolag, K., {et~al.} 2015, \apj, 812, 29

\bibitem[{Teklu {et~al.}(2017)Teklu, Remus, Dolag, \& Burkert}]{teklu+17}
Teklu, A.~F., Remus, R.-S., Dolag, K., \& Burkert, A. 2017, \mnras, 472, 4769

\bibitem[{Tenneti {et~al.}(2014)Tenneti, Mandelbaum, Di~Matteo, Feng, \& Khandai}]{tenneti+14}
Tenneti, A., Mandelbaum, R., Di~Matteo, T., Feng, Y., \& Khandai, N. 2014, \mnras, 441, 470

\bibitem[{Thater {et~al.}(2023)Thater, Jethwa, Lilley, Zocchi, Santucci, \& {van de Ven}}]{thater+23}
Thater, S., Jethwa, P., Lilley, E.~J., {et~al.} 2023, arXiv e-prints, arXiv2305.09344

\bibitem[{Thob {et~al.}(2019)Thob, Crain, McCarthy, Schaller, Lagos, Schaye, Talens, James, Theuns, \& Bower}]{thob+19}
Thob, A. C.~R., Crain, R.~A., McCarthy, I.~G., {et~al.} 2019, \mnras, 485, 972

\bibitem[{Tornatore {et~al.}(2007)Tornatore, Borgani, Dolag, \& Matteucci}]{tornatore+07}
Tornatore, L., Borgani, S., Dolag, K., \& Matteucci, F. 2007, \mnras, 382, 1050

\bibitem[{Tornatore {et~al.}(2004)Tornatore, Borgani, Matteucci, Recchi, \& Tozzi}]{tornatore+04}
Tornatore, L., Borgani, S., Matteucci, F., Recchi, S., \& Tozzi, P. 2004, \mnras, 349, L19

\bibitem[{Valenzuela \& Remus(2024)}]{valenzuela&remus24}
Valenzuela, L.~M. \& Remus, R.-S. 2024, \aap, 686, A182

\bibitem[{{Vallés-Pérez} {et~al.}(2020){Vallés-Pérez}, Planelles, \& Quilis}]{valles_perez+20}
{Vallés-Pérez}, D., Planelles, S., \& Quilis, V. 2020, \mnras, 499, 2303

\bibitem[{{van de Sande} {et~al.}(2019){van de Sande}, Lagos, Welker, {Bland-Hawthorn}, Schulze, Remus, Bahé, Brough, Bryant, Cortese, Croom, Devriendt, Dubois, Goodwin, Konstantopoulos, Lawrence, Medling, Pichon, Richards, Sanchez, Scott, \& Sweet}]{van_de_sande+19}
{van de Sande}, J., Lagos, C. D.~P., Welker, C., {et~al.} 2019, \mnras, 484, 869

\bibitem[{{van de Ven} {et~al.}(2008){van de Ven}, {de Zeeuw}, \& {van den Bosch}}]{van_de_ven+08}
{van de Ven}, G., {de Zeeuw}, P.~T., \& {van den Bosch}, R. C.~E. 2008, \mnras, 385, 614

\bibitem[{{van den Bosch} {et~al.}(2008){van den Bosch}, {van de Ven}, Verolme, Cappellari, \& {de Zeeuw}}]{van_den_bosch+08}
{van den Bosch}, R. C.~E., {van de Ven}, G., Verolme, E.~K., Cappellari, M., \& {de Zeeuw}, P.~T. 2008, \mnras, 385, 647

\bibitem[{{van der Marel} {et~al.}(1998){van der Marel}, Cretton, {de Zeeuw}, \& Rix}]{van_der_marel+98}
{van der Marel}, R.~P., Cretton, N., {de Zeeuw}, P.~T., \& Rix, H.-W. 1998, \apj, 493, 613

\bibitem[{Vasiliev \& Valluri(2020)}]{vasiliev&valluri20}
Vasiliev, E. \& Valluri, M. 2020, \apj, 889, 39

\bibitem[{Velliscig {et~al.}(2015)Velliscig, Cacciato, Schaye, Crain, Bower, {van Daalen}, Dalla~Vecchia, Frenk, Furlong, McCarthy, Schaller, \& Theuns}]{velliscig+15}
Velliscig, M., Cacciato, M., Schaye, J., {et~al.} 2015, \mnras, 453, 721

\bibitem[{{Vera-Ciro} {et~al.}(2011){Vera-Ciro}, Sales, Helmi, Frenk, Navarro, Springel, Vogelsberger, \& White}]{vera_ciro+11}
{Vera-Ciro}, C.~A., Sales, L.~V., Helmi, A., {et~al.} 2011, \mnras, 416, 1377

\bibitem[{Veršič {et~al.}(2024)Veršič, Rejkuba, Arnaboldi, Gerhard, Pulsoni, Valenzuela, Hartke, Watkins, {van de Ven}, \& Thater}]{versic+24}
Veršič, T., Rejkuba, M., Arnaboldi, M., {et~al.} 2024, arXiv e-prints, arXiv2403.13109

\bibitem[{Vincent \& Ryden(2005)}]{vincent&ryden05}
Vincent, R.~A. \& Ryden, B.~S. 2005, \apj, 623, 137

\bibitem[{Vogelsberger {et~al.}(2014)Vogelsberger, Genel, Springel, Torrey, Sijacki, Xu, Snyder, Nelson, \& Hernquist}]{vogelsberger+14:illustris}
Vogelsberger, M., Genel, S., Springel, V., {et~al.} 2014, \mnras, 444, 1518

\bibitem[{Wagner {et~al.}(1988)Wagner, Bender, \& Moellenhoff}]{wagner+88}
Wagner, S.~J., Bender, R., \& Moellenhoff, C. 1988, \aap, 195, L5

\bibitem[{Walker {et~al.}(2009)Walker, Mateo, Olszewski, Peñarrubia, Evans, \& Gilmore}]{walker+09}
Walker, M.~G., Mateo, M., Olszewski, E.~W., {et~al.} 2009, \apj, 704, 1274

\bibitem[{Weinberger {et~al.}(2017)Weinberger, Springel, Hernquist, Pillepich, Marinacci, Pakmor, Nelson, Genel, Vogelsberger, Naiman, \& Torrey}]{weinberger+17}
Weinberger, R., Springel, V., Hernquist, L., {et~al.} 2017, \mnras, 465, 3291

\bibitem[{Wiersma {et~al.}(2009)Wiersma, Schaye, \& Smith}]{wiersma+09}
Wiersma, R. P.~C., Schaye, J., \& Smith, B.~D. 2009, \mnras, 393, 99

\bibitem[{Wilkinson {et~al.}(2023)Wilkinson, Ludlow, Lagos, Fall, Schaye, \& Obreschkow}]{wilkinson+23}
Wilkinson, M.~J., Ludlow, A.~D., Lagos, C. d.~P., {et~al.} 2023, \mnras, 519, 5942

\bibitem[{Xu {et~al.}(2023{\natexlab{a}})Xu, Jing, \& Gao}]{xu+23}
Xu, K., Jing, Y.~P., \& Gao, H. 2023{\natexlab{a}}, \apj, 954, 2

\bibitem[{Xu {et~al.}(2023{\natexlab{b}})Xu, Jing, \& Zhao}]{xu+23:tng}
Xu, K., Jing, Y.~P., \& Zhao, D. 2023{\natexlab{b}}, \apj, 957, 45

\bibitem[{Zemp {et~al.}(2011)Zemp, Gnedin, Gnedin, \& Kravtsov}]{zemp+11}
Zemp, M., Gnedin, O.~Y., Gnedin, N.~Y., \& Kravtsov, A.~V. 2011, \apjs, 197, 30

\bibitem[{Zhang {et~al.}(2022)Zhang, Wuyts, Witten, Avery, Hao, Sharma, Shen, Toshikawa, \& Villforth}]{zhang+22}
Zhang, J., Wuyts, S., Witten, C., {et~al.} 2022, \mnras, 513, 4814

\end{thebibliography}

\begin{appendix}

\section{Testing the shape methods}
\label{app:shape_methods}

To test the typically used shape methods, we decided to apply the \emph{unweighted}, \emph{reduced}, and \emph{reduced ellipsoidal} methods to the galaxy sample used in this work. Given the definition of the second moment of the mass distribution tensor, $\tens{M}$ (\cref{eq:mass_tensor}), we let $\tilde{m}_k$ be the mass of particle $i$, and $w(\vec{r})$ is the weighting function that is different for the individual methods: $w(\vec{r}) = 1$ for unweighted, $w(\vec{r}) = r^{-2}$ for reduced, and $w(\vec{r}) = r_\mathrm{ell}^{-2}$ for reduced ellipsoidal, with $r_\mathrm{ell}$ defined as the ellipsoidal distance in \cref{eq:rell}. While the latter can only be applied iteratively because of the dependence of $r_\mathrm{ell}$ on the ellipsoid axis ratios, the unweighted and reduced methods can also be applied directly without iteration. For the iterative procedures, there are two possibilities of modifying the ellipsoid, (1) by keeping the major axis of the ellipsoid constant, or (2) by keeping the volume constant.

\begin{figure*}
    \centering
    \includegraphics[width=0.8\textwidth]{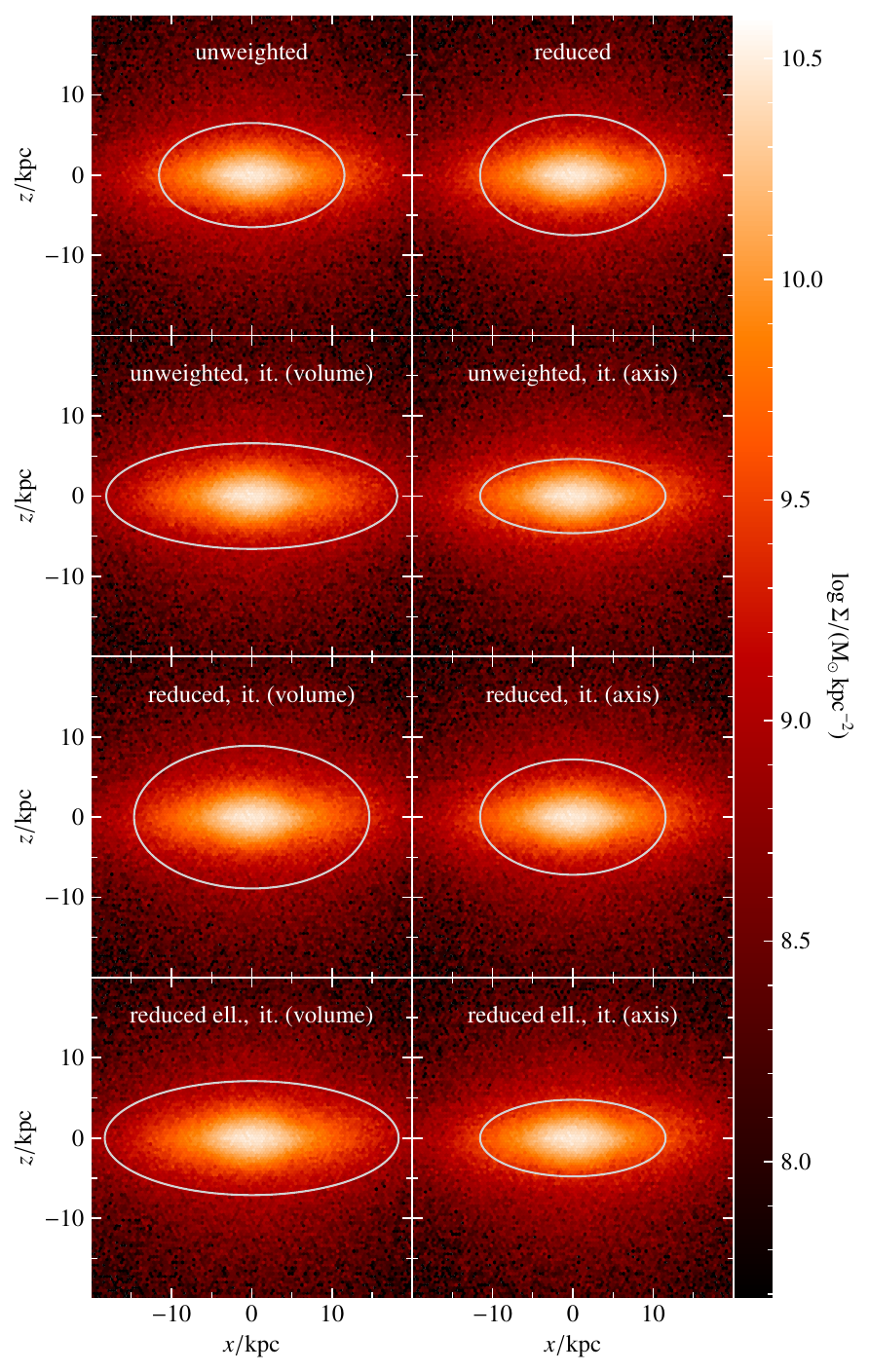}
    \caption{Surface density map of an example galaxy with the overplotted ellipsoid contour at \SI{1}{\Rhalf} for the considered shape determination methods. The galaxy is viewed edge-on in the eigenvector coordinate system of the respective shape tensors. The iterative methods are denoted by \enquote{it.}.}
    \label{fig:density_map_shape_methods}
\end{figure*}

In total, this results in eight different shape determination methods. \Cref{fig:density_map_shape_methods} shows all of these applied to the stellar component of one example galaxy of Magneticum Box4 (uhr) at an initial radius of \SI{1}{\Rhalf} and then rotated to the edge-on perspective based on the shape ellipsoid. The elliptical contours of the ellipsoids are also overdrawn on the density maps to visualize how well the shape determination methods match the expected shape contours. As commented by \citet{zemp+11}, the physical meaning of the shapes obtained from the non-iterative methods is unclear. Applying the unweighted and reduced methods to the galaxies results in a strong bias towards spherical shapes, which can clearly be seen in the first row of \cref{fig:density_map_shape_methods}. The reason for this is that considering all the particles well above and below the disk plane skews the resulting mass tensor to the initial shape of the volume in which the particles are considered. Some have attempted to correct this bias by modifying the computed axis ratios \citep[e.g.,][]{bailin&steinmetz05,knebe+10}. However, only by iteratively deforming the volume is this bias reduced until convergence.

\begin{figure}
    \centering
    \includegraphics[width=\columnwidth]{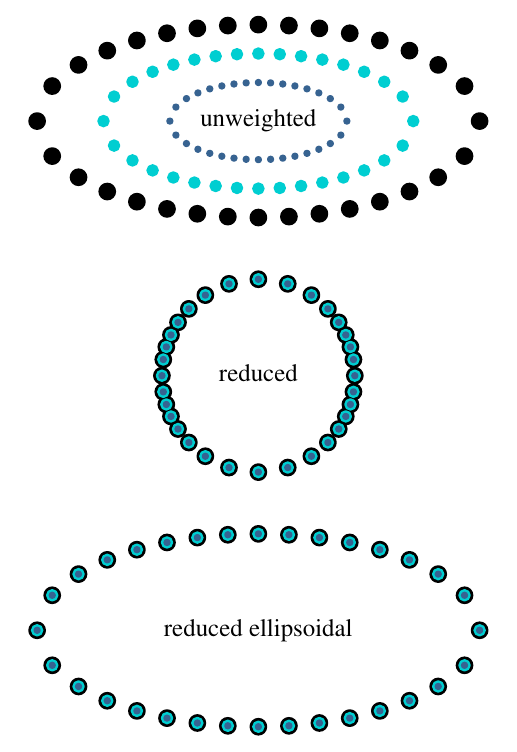}
    \caption{Visualization in two dimensions of how the unweighted, the reduced, and the reduced ellipsoidal weight functions project the particle distribution onto itself, onto a sphere (circle in 2D), and onto an ellipsoid (ellipse in 2D) in the summation for the shape tensor. The colors and sizes of the particles only are different to emphasize that the reduced and reduced ellipsoidal weightings project their positions onto the same circle and ellipsoid, respectively.}
    \label{fig:reduced_visualization}
\end{figure}

It also becomes apparent that the iterative reduced method leads to shapes that are too spherical independent of the quantity kept constant throughout the iteration. The reason for this is related to the weighting $w(\vec{r}) = r^{-2}$ of the individual summands for the mass tensor. This weighting leads to all particles being treated as though they were lying on a unit sphere around the center of the galaxy, with their positions on the sphere given by their normalized position vectors. Because of the spherical arrangement of particles, this leads to a bias towards spherical shapes. A 2D visualization is shown in \cref{fig:reduced_visualization}, which also shows how the unweighted method treats all particles at their original positions because of its weighting of $w(\vec{r}) = 1$. In contrast, the reduced ellipsoidal method treats all particles as though they were located on a unit ellipsoid with the axis ratios of the converging shape because of the weighting of $w(\vec{r}) = r_\mathrm{ell}^{-2}$. This illustrates why neither the unweighted nor the reduced ellipsoidal methods have a bias towards spherical shapes. As \citet{zemp+11} discuss, the axes obtained from the unweighted and reduced ellipsoidal methods are expected to give the same results for ellipsoidal shells, whereas the interpretation of the axes for the reduced method is unclear.

The difference between the unweighted and reduced ellipsoidal methods therefore lies in the projection onto a unit ellipsoid or the lack thereof. Projecting all particles onto the same surface essentially means that an average shape over all particles enclosed in the volume is taken. The advantage of this method is that all particles are equally weighted \citep[e.g.,][]{dubinski&carlberg91}, which limits the influence of substructure towards the edge of the considered volume. Since we remove such substructure through the use of the halo finder \subfind{} and are interested in the actual local shape at a given radius instead of an average over all particles within the volume, we decided to use the iterative unweighted method. As \citet{zemp+11} showed, this method also leads to more accurate local shapes than the reduced ellipsoidal method, which is too strongly influenced by the shape behavior in the inner region of the considered volume.

Finally, the choice of the ellipsoidal quantity kept constant during the iteration affects the size of the final shape ellipsoid. For near-spherical shapes, there is very little difference between the two. Keeping the major axis constant results in smaller volumes the flatter the converged shape is. Therefore, for the same starting radius, the mass in more spherical shapes will generally be larger than in flattened ones. When comparing galaxy shapes at the same radius, however, we decided that it is more appropriate to compare the shapes of ellipsoids with equal volumes instead of different ones that only have the same major axis length. The iteration method with the constant major axis length is additionally less stable than when keeping the volume constant. In rare cases, in which a large overdensity of particles lies towards the border of the starting radius, the overdensity can lead to the shape flattening entirely to a plane for a constant major axis length. For example, this can happen in the case of an ongoing merger not being identified as two distinct subhalos by the halo finder. Of course, it should be noted that an ellipsoidal shape is no longer an appropriate measure of the shape in such cases. Because of the two stated reasons, for this work we decided to perform the shape determinations with the iterative unweighted method with a constant ellipsoidal volume throughout the iterations.

\section{Complementary shape relations with the stellar component}
\label{app:stellar_global_properties}

\begin{figure*}
    \centering
    \includegraphics{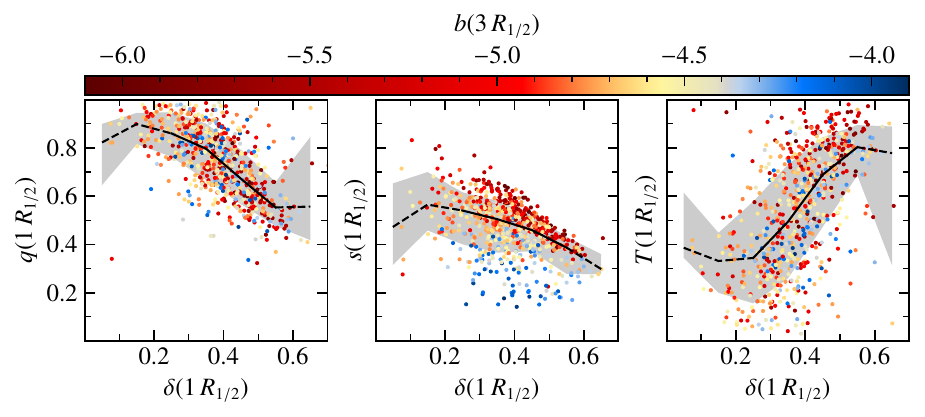}
    \includegraphics{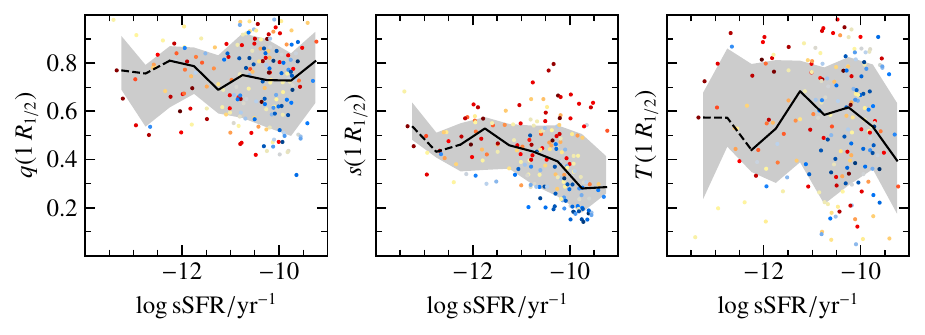}
    \includegraphics{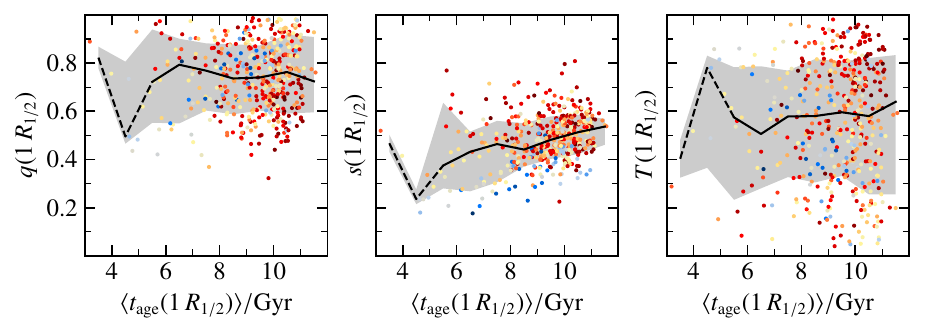}
    \caption{Relation between shape parameters $q$, $s$, and $T$ at one stellar half-mass radius and different stellar properties found within the half-mass radius, colored by the \bvalue{}.
    \emph{Top row}: Anisotropy of the stellar component within $\SI{1}{\Rhalf}$.
    \emph{Middle row}: Specific star formation rate within $\SI{1}{\Rhalf}$. Only galaxies with non-zero star formation in their centers are plotted.
    \emph{Bottom row}: Mass-weighted average age of the stars within $\SI{1}{\Rhalf}$.
    The black lines indicate the median values in the respective bins and the shaded regions the $1\sigma$ ranges (containing \SI{68}{\percent} of the galaxies above and below the median).}
    \label{fig:stellar_global_properties_appendix}
\end{figure*}

As the stellar shapes at \SI{1}{\Rhalf} are correlated with the kinematic quantity \lambdaR{} (\cref{fig:lambda_r_bvalue}), this also means that they are related to the anisotropy $\delta$ \citep{illingworth77,binney78,binney05,emsellem+11:atlas3dIII,schulze+18}.
We took the values of the anisotropy obtained for our simulated galaxy sample by \citet{schulze+18}, who used the following definition:
\begin{equation}
    \delta = 1 - \frac{\Pi_{zz}}{\Pi_{xx}} = 1 - \frac{\sum_{i=1}^N M_i \sigma_{z,i}^2}{\sum_{i=1}^N M_i \sigma_{x,i}^2},
\end{equation}
where the sums run over 3D cubic cells into which the particles were binned, $M_i$ is the mass of a bin cell, and $\sigma_{x,i}$ and $\sigma_{z,i}$ are the cell's velocity dispersions in $x$- and $z$-direction, respectively.
The relations between $\delta$ and the three shape parameters $q$, $s$, and $T$, all measured within and at \SI{1}{\Rhalf}, are shown in the top row of \cref{fig:stellar_global_properties_appendix}. All three relations show clear trends, and especially the edge-on axis ratio $s$ is tightly correlated with the anisotropy when accounting for the different morphological types.

In \cref{sec:stellar_shapes}, we found a strong relation between the edge-on axis ratio $s$ and the galaxy morphology quantified by the \bvalue{}. As LTGs tend to be more star-forming than ETGs, we expected there to also be a relation between $s$ and the specific star formation rate ($\mathrm{sSFR}$), and also with the mean central stellar age. Both relations are shown in the bottom two rows of \cref{fig:stellar_global_properties_appendix}, for which we determined $\mathrm{sSFR}$ and the mass-weighted mean central stellar age within \SI{1}{\Rhalf}. Galaxies with higher specific star formation rate tend to be LTGs and are therefore flatter (lower values of $s$), whereas older galaxies are generally more spherical (higher values of $s$), just as expected from the fact that we found LTGs to be flatter.
Finally, just as the correlation between the morphology and $q$ and $T$ is weaker, there are no significant trends found for $q$ and $T$ with the specific star formation rate or the mean stellar age.

\section{Angular momenta}
\label{app:angular_momenta}

\begin{figure}
    \centering
    \includegraphics{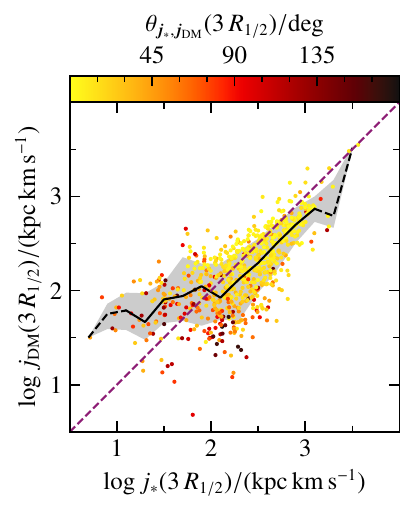}
    \caption{
    Relation between the absolute values of the stellar and DM angular momenta, colored by the angle between their vectors, $\theta_{\jstvec, \jdmvec}$, for within three half-mass radii. The dashed blackberry line denotes where the absolute angular momenta are equal to each other.
    }
    \label{fig:angular_momentum_introduction}
\end{figure}

The absolute stellar and DM angular momenta ($j_*$ and $j_\mathrm{DM}$, respectively) within \SI{3}{\Rhalf} are related to each other, particularly for larger values (see \cref{fig:angular_momentum_introduction}). Around \SI{75}{\percent} of the galaxies have a larger stellar angular momentum than the DM has. For values of $j_* \gtrsim \SI{e2}{\kilo\parsec\kilo\meter\per\second}$, we found a rising lower bound for $j_\mathrm{DM}$, which is accompanied by a strong alignment of the angular momentum vectors. In contrast, galaxies with low stellar angular momentum tend to have large misalignments of the stellar and DM angular momenta, the correlation of the absolute values breaks down, and they become largely independent of each other. This is related to the larger uncertainties of the orientation of the angular momentum the smaller its absolute value is. The almost constant median DM angular momentum for low values of $j_*$ indicates that the DM is largely decoupled from the stellar component for low angular momentum galaxies. As can be seen in \cref{fig:alignment_histogram}, poor alignment between the angular momenta are typically found in ETGs, suggesting that such configurations are likely linked to violent merger histories.

\end{appendix}

\end{document}